\documentclass[%
 aip,
 amsmath,amssymb,
reprint, onecolumn 
]{revtex4-1}

\usepackage{graphicx}
\usepackage{dcolumn}
\usepackage{bm}

\usepackage[utf8]{inputenc}
\usepackage[T1]{fontenc}
\usepackage{mathptmx}
\usepackage{etoolbox}

\usepackage{eurosym}
\usepackage{graphicx} 
\usepackage{amsfonts}
\usepackage{amsmath}
\usepackage{amsfonts}
\usepackage{amssymb}
\usepackage{amsmath}
\usepackage{epsfig}
\usepackage{graphicx}
\usepackage{xcolor}
\usepackage{afterpage}
\usepackage{bigints}
\usepackage{fancyhdr}
\usepackage{subcaption}
\usepackage{hyperref}
\usepackage{multirow}
\usepackage{array}
\usepackage{booktabs}

\makeatletter
\def\@email#1#2{%
 \endgroup
 \patchcmd{\titleblock@produce}
  {\frontmatter@RRAPformat}
  {\frontmatter@RRAPformat{\produce@RRAP{*#1\href{mailto:#2}{#2}}}\frontmatter@RRAPformat}
  {}{}
}%
\makeatother
\begin{document}

\preprint{AIP/123-QED}

\title[Hele-Shaw Flow With Pressure and Shear Rate Dependent Viscosity]{Hele-Shaw Flow With Pressure and Shear Rate Dependent Viscosity}
\author{Benedetta Calusi}
 \affiliation{Dipartimento di Matematica e Informatica "Ulisse Dini", Universita degli Studi di Firenze, Viale Morgagni 67/a, 50134 Firenze, Italy}

\author{Liviu Iulian Palade}%
 \email{benedetta.calusi@unifi.it}
 \email{liviu-iulian.palade@insa-lyon.fr}
\affiliation{ 
Institut Camille Jordan CNRS UMR5208, INSA-Lyon \& P\^ole de Math\'ematiques, Universit\'e de Lyon, 21 Avenue Jean Capelle, 69621 Villeurbanne Cedex, France
}%


\date{\today}

\begin{abstract}
This paper investigates the behavior of a fluid characterized by a viscosity simultaneously depending  on pressure and shear rate within a Hele-Shaw cell featuring a sharp corner geometry. The study extends previous analyses conducted on purely pressure-dependent (piezo-viscous) and yield-stress fluids, providing a new perspective on confined complex flows. Motivated by practical applications related to designing biomedical devices and flows of relevance to bio-medicine area,  thin film technologies, injection molding - to name only a few - the flow configuration considered here can highlight essential features of complex fluid behavior in narrow-gap geometries around a sharp edge. Starting from the governing equations for an incompressible generalized Newtonian fluid and employing an appropriate rheological model, we derive the modified flow equations adapted to the Hele-Shaw flow. A particular solution is obtained near the corner region. Numerical simulations complement the theoretical results, illustrating the influence of the rheological parameters on the flow behavior. 
\end{abstract}

\maketitle

\section{Introduction}\label{int}

Flows between two narrowly spaced, parallel plates - commonly referred to as \emph{Hele-Shaw flows} - represent a classical configuration in fluid mechanics, originally introduced in the pioneering work by Hele-Shaw~\cite{HELESHAW1898}. Owing to their mathematical tractability and physical relevance, such flows have received considerable attention over the years. They serve as a prototypical model for a variety of complex flow phenomena and provide valuable insights into the behavior of fluids in confined geometries. General overviews and comprehensive treatments of Hele-Shaw flows, including discussions on instabilities and pattern formation, yield stress fluids and temperature effects upon, can be found in~\cite{jjx1998, pop2001, nb2006, coussot2014,  huilgeo2015, pop2017, evans2025}.

The Hele-Shaw configuration is particularly well-suited to studying flows past obstacles of different geometries. These obstacles, which may be embedded between parallel surfaces (e.g. plates), can significantly alter the flow field and introduce complex local phenomena. Cylindrical inclusions orthogonal to the plates have been widely studied~\cite{Hassager1988-xz}, while more intricate geometrical features, such as sharp corners and edges, have also drawn interest due to their relevance in engineering and natural systems~\cite{Chupin2008, Hassager1988-xz}. In such cases, singularities or boundary layers may develop near the geometric discontinuities, necessitating refined mathematical treatment.

Depending on the application, the plates may be aligned horizontally or inclined, and the flow may involve various classes of fluids ranging from simple Newtonian  to complex non-Newtonian fluids (such as viscoplastic, yield stress  materials). Hele-Shaw flows are encountered in a wide spectrum of scientific and industrial contexts, including biological systems and biomedicine, microfluidics, injection molding, thin-film lubrication, porous media transport.  Applications include, and are not limited to, biomedical devices, tribology, inkjet printing, geological flows. Numerous studies have addressed complex fluids flow in different problem geometry~\cite{Allouche2015, Allouche2017, BALMFORTH2004, Benjamin1957, borsi-2008,  Chakraborty2019, Farina2018, Gholinezhad2023-dh, Hintermuller2021, Li2022, taylor2022, Pascal_1999, Petit2024-fi, Rajagopal2012, Falsaperla2020, Fusi2018}.

The theoretical foundation for modeling complex flows often relies on simplifications derived from the narrow-gap assumption, leading to depth-averaged equations. Nonetheless, considerable complexity remains, especially when dealing with fluids whose rheological properties may vary with pressure or shear rate. For instance, the use of viscometric function formulations in~\cite{Huilgol2006} allows for a unified treatment of viscous and viscoplastic behaviors under Hele-Shaw conditions. Recent contributions have further explored experimental observations, numerical modeling, interface instabilities such as viscous fingering, and boundary layer effects near geometrical singularities~\cite{Kislaya2025, Delmastro2025, Krakov2025, Zahid2025, Firoozi2025, Delmastro2024, Daripa2024}. Moreover, recent developments have introduced extensions and generalizations of Hele-Shaw models, including anisotropic effects, surface tension influences, and multi-phase flows~\cite{Ouyang2024}.

\begin{figure}
    \centering
    \includegraphics[width=0.7\linewidth]{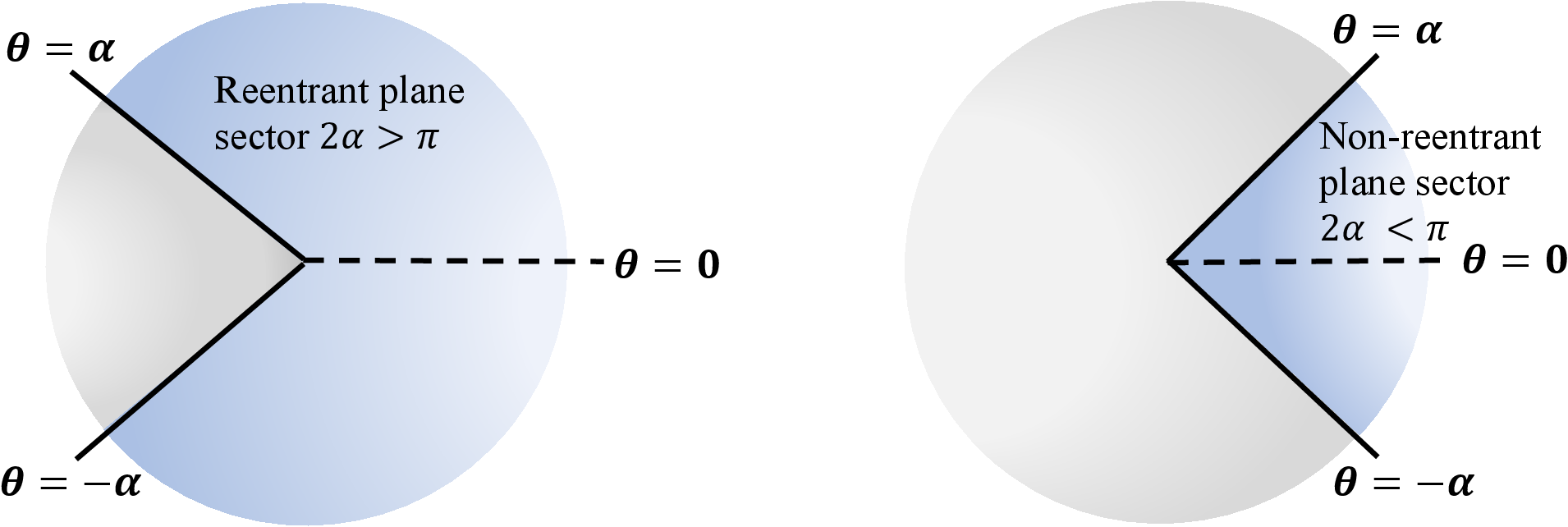}
    \caption{Illustration of the problem geometry for different edge angles, showing a sharp edge configuration (reentrant sector) and a cavity-like configuration (non-reentrant sector) near the tip region.}
    \label{fig:angles}
\end{figure}

The present paper is concerned with Hele-Shaw flows past sharp corners (see Figure \ref{fig:angles}) and aims to extend previous studies by incorporating more complex rheological behavior. In particular, we build upon the works~\cite{Chupin2008} and~\cite{Calusi2024b}. The former investigated two-dimensional symmetric and anti-symmetric flows of generalized Newtonian and Herschel-Bulkley fluids, while the latter considered a piezo-viscous fluid, where the viscosity depends on pressure. Here, we advance the analysis by considering a \emph{piezo-shear fluid} - a fluid whose viscosity is a function of both pressure and shear rate.  To the best of our knowledge, this combined dependence has not yet been studied in the context of Hele-Shaw corner flows.  The modeling of various industrial flows wherein the range of pressures is large, e.g. injection molding of complex fluids, flows of relevance to elastohydrodynamics - to cite only a few -  requires that the effect of pressure on the
viscosity must be accounted for.

Understanding the interaction between pressure- and shear-dependent viscosity in confined geometries with sharp boundaries is of both theoretical and practical importance. It may lead to more accurate predictions of flow behavior in technological processes and provides new insights into the interplay between geometry and complex fluid rheology.

The paper is organized as follows. In Section~\ref{mb}, we introduce the governing equations and the constitutive model for the generalized piezo-shear fluid. Section~\ref{ps} presents an asymptotic analysis of the flow near the sharp corner, highlighting the singular behavior and its dependence on the rheological parameters.  Section~\ref{dc} is devoted to numerical results and their interpretation, and the final conclusions  are gathered in Section ~\ref{cn}.

\section{Mathematical Background}\label{mb}

  In the following the notation  ``$*$" denotes dimensional quantities.  We consider - in a way similar in nature to the work by Rajagopal, Saccomandi and Vergori in~\cite{Rajagopal2012} - a fluid with pressure $p^*$ and shear rate $\Dot{\gamma}^*$  dependent viscosity $\mu^*$ as defined below:


\begin{equation}\label{visc}
 \mu^* ( p^*
, \Dot{\gamma}^*) = f_1^*(p^*)f_2^*(\Dot{\gamma}^*) = f_1^*(p^*) \vert\vert \boldsymbol{A}_1\vert\vert ^{\frac{1-\lambda}{\lambda}} , \quad f_2^*(\Dot{\gamma}^*) = \vert\vert \boldsymbol{A}_1\vert\vert ^{\frac{1-\lambda}{\lambda}}   ,   
\end{equation}

where $f_1^*(p^*)$ is the pressure dependent function that is defined in \eqref{ceq} below, $\lambda > 0$ is a dimensionless parameter controlling the degree of nonlinearity in the shear response, and  
$\boldsymbol{A}_1^*$ denotes the (traceless) first Rivlin--Ericksen tensor:

\begin{equation}
    \boldsymbol{A}_1^* = \dfrac{1}{2} \left[ \nabla \boldsymbol{u}^* + \left( \nabla \boldsymbol{u}^* \right)^T \right] , 
\end{equation}

with $\vert\vert \boldsymbol{A}_1\vert\vert $ being its norm and $ \boldsymbol{u}^*  $  the velocity field.  It has the advantage of being sufficiently versatile as it is able to describe the viscosity patterns of a quite large family of fluids.  Specifically,
the parameter $\lambda$ characterizes the fluid’s shear-dependent behavior in the following way: shear-thinning behavior is exhibited for $\lambda > 1$, and shear-thickening behavior for $\lambda \in (0,1)$.   It is common to introduce the viscosity equation in the form of a product  $f_1^*(p^*)f_2^*(\Dot{\gamma}^*)$, it also has the advantage to ease subsequent calculations.

In the introductory section of~\cite{Rajagopal2012} a detailed discussion about the viscosity dependence upon pressure is also given and we shall gather inspiration from it for the following clarifying observations:  

\begin{itemize}

\item the Authors of~\cite{Rajagopal2012}  mentioned that it has experimentally been observed that the change in the density of an organic fluid subjected to very large pressure range is about 3--5\% while the change in the viscosity is of order $10^8$\%.  We therefore concur with these Authors that based on the experimental evidence cited therein most organic fluids can reasonably be considered as incompressible (as    is here stated in equation \eqref{Sys:GenGovEq-0}).

\item regarding the meaning of $p^*$ in the viscosity constitutive equation and its relation to the hydrodynamic pressure: in~\cite{Rajagopal2012}  it is stated that (and we quote) "the mean normal stress in an incompressible nonlinear fluid need not be the  `pressure' in the fluid, if by  `pressure' we mean the reaction stress due to the constraint of incompressibility".  As in~\cite{Rajagopal2012} we consider the fluid viscosity to depend on the mean normal stress (the pressure).  
\end{itemize}


Let now the fluid constitutive equation be given by

\begin{equation}\label{ceq}
    \boldsymbol{\mathbb{S} }^* = 2 \mu ( p^*
, \Dot{\gamma}^*) \boldsymbol{A}_1^*
= 2 f_1^*(p^*)f_2^*( \Dot{\gamma}^* )\boldsymbol{A}_1^*   . 
\end{equation}
where $\boldsymbol{\mathbb{S} }^*$ denotes the Cauchy extra-stress tensor.

 \begin{figure}[!h]
    \centering
    \includegraphics[width=0.6\linewidth]{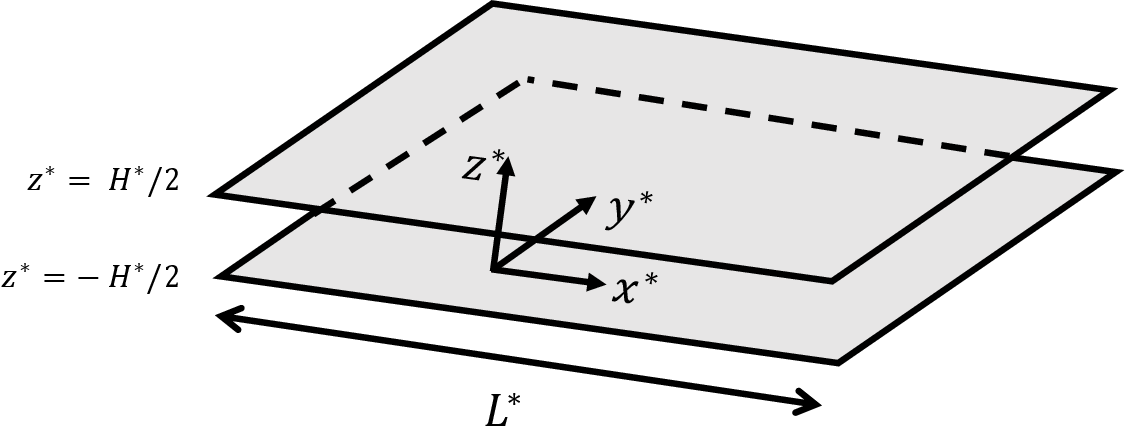}
    \caption{Overview of the geometric configuration.}
    \label{fig:schematic}
\end{figure}

The (momentum balance) governing equations for the fluid motion are 

\begin{equation}
    \begin{cases}
        \rho^* \left( \dfrac{\partial \boldsymbol{u}^*}{\partial t^*} + \boldsymbol{u}^*\cdot \nabla \boldsymbol{u}^*\right) = 
     - \nabla p^* + \mathrm{div} \boldsymbol{\mathbb{S}}^* , \\
     \\
     \mathrm{div} \boldsymbol{u}^* = 0 , 
    \end{cases}\label{Sys:GenGovEq-0}
\end{equation}

where $\rho^*$ is the constant fluid density. Figure \ref{fig:schematic} shows the schematic diagram of the problem domain. 

Let the velocity field be of the form 
\begin{equation}
    \boldsymbol{u}^*(x^*,y^*,z^*) = u^*(x^*,y^*,z^*) \boldsymbol{e}_x^* + v^*(x^*,y^*,z^*) \boldsymbol{e}_y^* +  w^*(x^*,y^*,z^*) \boldsymbol{e}_z^* , 
    \label{Def:u-v-w}
\end{equation}
and the gap between the horizontal $x^*0y^*$ plane and any of the two boundary planes be denoted $\vert z^*\vert = H^*/2$. 

We consider the following boundary conditions 
\begin{equation}
    \boldsymbol{u}^* = 0 , \qquad \mathrm{for}\ \  z^* = \pm \dfrac{H^*}{2}, \label{BC:NoWallSlip-0}
\end{equation}
i.e., the no wall slip condition, and, due to the flow domain symmetry 
\begin{equation}
    \dfrac{\partial u^*}{\partial z^*} = \dfrac{\partial v^*}{\partial z^*} = 0 , \qquad \mathrm{for}\ \ z^* = 0, \label{BC:FlowDomSym-0}
\end{equation}
i.e., we focus only on the upper half subdomain $ z^* \in \left[ 0,H^*/2 \right] $. 
For the Hele-Shaw flows the fluid film thickness is negligible compared to the domain $x^*-y^*$ other dimensions. Since the flow is chiefly confined to the $x^*$ and $y^*$ directions, it is usually assumed that $w^*$ is negligible compared to $u^*$ and $v^*$. 

We introduce the following dimensionless quantities
\begin{equation}
\begin{array}{c}
x=\dfrac{x^{\ast }}{L^{\ast }},\quad y=\dfrac{y^{\ast }}{L^{\ast }}, \quad z=\dfrac{z^{\ast }}{H^{\ast }}, \quad 
u=\dfrac{{u}^{\ast }}{U_{ref}^{\ast }},\quad v=\dfrac{{v}^{\ast }}{U_{ref}^{\ast }}, \quad w=\dfrac{{w}^{\ast }}{\epsilon U_{ref}^{\ast }}, 

 \\
\\
t={\displaystyle{\frac{U_{ref}^*}{L^{\ast }}}}t^{\ast }, \quad  p=\dfrac{p^{\ast }-p_a^*}{p_c^*},\quad 
\mathbb{S} = 
\dfrac{\mathbb{S}^{\ast }}{S_c^*},\ \ \ f_1 (p) = \dfrac{\mu_{c}^{\ast }(p^*)}{ \mu^{\ast }_c } , 
\end{array}
\label{adim1}
\end{equation}%
with  
\begin{equation}
    \epsilon = \dfrac{H^*}{L^*} \ll 1 , 
\end{equation}
\begin{equation}
    p_c^* = \dfrac{ \mu_{c}^{\ast }U_{ref}^{\ast } L^* }{ H^{\ast^2 }
} , \qquad S_c^* = \dfrac{ \mu_{c}^{\ast }U_{ref}^{\ast } }{ H^{\ast}} ,
\end{equation}
where $U_{ref}^*$ is the reference velocity, $p_a^*$ is the reference (atmospheric) pressure and $\mu_c^*$ is the characteristic fluid viscosity. 

Let the Reynolds number be defined as follows:

\begin{equation}\label{ReNumber}
\mathsf{Re}={\displaystyle{\frac{\rho ^{\ast
}U_{ref}^{\ast }H^{\ast }}{\mu_{c}^{\ast }}}}.
\end{equation}

As this is a first step towards a more realistic modeling, we do not consider in this paper a pressure dependent (via the viscosity) Reynolds number.

Next, using \eqref{Def:u-v-w} and the
adimensionalization \eqref{adim1}, system \eqref{Sys:GenGovEq-0}  rewrites 
\begin{equation}
    \begin{cases}
        \epsilon \mathsf{Re} \left( u\dfrac{\partial u}{\partial x} + v\dfrac{\partial u}{\partial y} + w\dfrac{\partial u}{\partial z}\right) = - \dfrac{\partial p}{\partial x} + 
         \biggl\{ 2\epsilon^2 \dfrac{\partial }{\partial x }\left( f_1 (p) f_2(\Dot{\gamma}) \dfrac{\partial u}{\partial x} \right)  \\ 
          \qquad\qquad\qquad + \epsilon^2 \dfrac{\partial }{\partial y }\left[ f_1 (p) f_2(\Dot{\gamma})  \left( \dfrac{\partial v}{\partial x} + \dfrac{\partial u}{\partial y}\right)\right] + \dfrac{\partial }{\partial z }\left[ f_1 (p) f_2(\Dot{\gamma}) \left(\epsilon^2 \dfrac{\partial  w}{\partial x }  + \dfrac{\partial  u}{\partial z}\right) \right]\biggl\}, \\
        \\
        \epsilon \mathsf{Re} \left( u\dfrac{\partial v}{\partial x} + v\dfrac{\partial v}{\partial y} + w\dfrac{\partial v}{\partial z}\right) = - \dfrac{\partial p}{\partial y} + 
         \epsilon^2\dfrac{\partial }{\partial x }\left[ f_1 (p) f_2(\Dot{\gamma}) \left( \dfrac{\partial v}{\partial x} + \dfrac{\partial u}{\partial y}\right)\right]   
         \\ 
         \qquad\qquad\qquad + 2\epsilon^2 \dfrac{\partial }{\partial y }\left( f_1 (p) f_2(\Dot{\gamma})  \dfrac{\partial v}{\partial y} \right)+ \dfrac{\partial }{\partial z }\left[ f_1 (p) f_2(\Dot{\gamma}) \left( \dfrac{\partial v}{\partial z} + \epsilon^2 \dfrac{\partial w}{\partial x}\right)\right] , \\    
        \\
        \epsilon^2 \mathsf{Re} \left( u\dfrac{\partial w}{\partial x} + v\dfrac{\partial w}{\partial y} + w\dfrac{\partial w}{\partial z}\right) = - \dfrac{\partial p}{\partial z} + 
        \dfrac{\partial }{\partial x }\left[ f_1 (p) f_2(\Dot{\gamma}) \left(  \epsilon \dfrac{\partial u}{\partial z } + \epsilon^3 \dfrac{\partial w}{\partial x } \right)\right]  \\ 
        \qquad\qquad\qquad  + \dfrac{\partial }{\partial y }\left[ f_1 (p) f_2(\Dot{\gamma}) \left( \epsilon \dfrac{\partial v}{\partial z } + \epsilon^3\dfrac{\partial w}{\partial y } \right)\right]+ 2\epsilon  \dfrac{\partial }{\partial z }\left( f_1 (p) f_2(\Dot{\gamma}) \dfrac{\partial w}{\partial z }\right),   \label{Sys:GenGovEq-adim}        
    \end{cases}
\end{equation}
where 
\begin{multline}\label{sys2}
    f_1 (p) f_2(\Dot{\gamma}) = f_1 (p) \; \Dot{\gamma}^{\frac{1-\lambda}{\lambda}} = f_1 (p) \left[  \epsilon^2 \left(  \dfrac{\partial u}{\partial x }\right)^2 + \epsilon^2 \left(  \dfrac{\partial v}{\partial y }\right)^2 + \epsilon^2 \left(  \dfrac{\partial w}{\partial z}\right)^2   +  \dfrac{\epsilon^2}{2} \left(  \dfrac{\partial u}{\partial y } + \dfrac{\partial v}{\partial x }\right)^2 \right. \\ 
    \left. +  \dfrac{1}{2} \left(  \dfrac{\partial u}{\partial z }   + \epsilon^2 \dfrac{\partial w}{\partial x }\right)^2 + \dfrac{1}{2} \left(  \dfrac{\partial v}{\partial z } + \epsilon^2 \dfrac{\partial w}{\partial y }\right)^2  \right]^{\frac{1-\lambda}{2\lambda}}. 
\end{multline}
By assuming $\mathsf{Re} = \mathcal{O}(1)$ and neglecting terms of order $\epsilon$ and higher, system \eqref{Sys:GenGovEq-adim} reduces to 
\begin{equation}
    \begin{cases}
        \dfrac{\partial p}{\partial x} =  \dfrac{\partial }{\partial z} \left( \mu(p,\Dot{\gamma}) \dfrac{\partial u}{\partial z}  \right) , \\
        \\
        \dfrac{\partial p}{\partial y} =   \dfrac{\partial }{\partial z} \left( \mu(p,\Dot{\gamma})  \dfrac{\partial v}{\partial z}  \right)  , \\        
        \\ 
        \dfrac{\partial p}{\partial z} = 0 , \\
        \\
        \dfrac{\partial u}{\partial x} + \dfrac{\partial v}{\partial y} + \dfrac{\partial w}{\partial z} = 0,  
    \end{cases}\label{Sys:RedGovEq0}
\end{equation}

where, recalling formula \eqref{sys2}, $\mu(p,\Dot{\gamma})$ reduces to 
\begin{equation}
        \mu(p,\Dot{\gamma})  =  f_1 (p) f_2(\Dot{\gamma}) =   f_1 (p) \; \Dot{\gamma}^{\frac{1-\lambda}{\lambda}} 
        =   f_1 (p) \left[  \dfrac{1}{2}\left(  \dfrac{\partial u}{\partial z }\right)^2  + \dfrac{1}{2}  \left(  \dfrac{\partial v}{\partial z } \right)^2  \right]^{\frac{1-\lambda}{2\lambda}} =  2^{-\frac{1-\lambda}{2\lambda}}  f_1 (p) \left[  \left(  \dfrac{\partial u}{\partial z }\right)^2  +   \left(  \dfrac{\partial v}{\partial z } \right)^2  \right]^{\frac{1-\lambda}{2\lambda}}. 
\end{equation}
System \eqref{Sys:RedGovEq0} is to be coupled with boundary conditions \eqref{BC:FlowDomSym-0} which in dimensionless formulation are 
\begin{equation}
    \boldsymbol{u} = 0 , \qquad \mathrm{for}\ \ z =  \dfrac{1}{2}, \label{BC:NoWallSlip}
\end{equation}
and 
\begin{equation}
    \dfrac{\partial u}{\partial z} = \dfrac{\partial v}{\partial z} = 0 , \qquad \mathrm{for}\ \ z = 0. \label{BC:FlowDomSym}
\end{equation}
Also system \eqref{Sys:RedGovEq0} entails 
\begin{equation}
    p=p(x,y),
\end{equation}
and, thus, equations \eqref{Sys:RedGovEq0}$_{1,2}$ become 
\begin{equation}
    \begin{cases}
        \dfrac{\partial p}{\partial x} = 2^{-\frac{1-\lambda}{2\lambda}} f_1(p)\dfrac{\partial }{\partial z} \left(  f_2(\Dot{\gamma}) \dfrac{\partial u}{\partial z}  \right)   , \\
        \\
        \dfrac{\partial p}{\partial y} = 2^{-\frac{1-\lambda}{2\lambda}} f_1(p) \dfrac{\partial }{\partial z} \left(  f_2(\Dot{\gamma}) \dfrac{\partial v}{\partial z}  \right)    . 
    \end{cases}\label{Sys:RedGovEq}
\end{equation}
By proceeding as in \cite{Chupin2008}, we get
\begin{equation}
  \Dot{\gamma} = 
  \left(2^{\frac{1-\lambda}{2\lambda}}\dfrac{ zX(\vert\vert p\vert\vert ) }{f_1(p)}\right)^\lambda, \qquad X(\vert\vert p\vert\vert ) = \sqrt{\left(   \dfrac{\partial p}{\partial x}\right)^2 + \left(   \dfrac{\partial p}{\partial y}\right)^2} ,
\end{equation}
and, thus, 
the velocity field is given by 
\begin{equation}
    \begin{cases}
        u(x,y,z) = \displaystyle\int_{H/2}^z \dfrac{\partial u}{\partial z} dt = 
        \dfrac{1}{X(\vert\vert p\vert\vert)} \left( \displaystyle{\int}_{H/2}^z \left(2^{\frac{1-\lambda}{2\lambda}}\dfrac{ t\,X(\vert\vert p\vert\vert ) }{f_1(p)}\right)^\lambda dt\right) \dfrac{\partial p}{\partial x}(x,y)
        \\ 
        \qquad\qquad\qquad\qquad\quad = \dfrac{ 2^{\frac{1-\lambda}{2}} \left( X(\vert\vert p\vert\vert )\right)^{\lambda-1}}{\left(f_1(p)\right)^\lambda} \dfrac{\partial p}{\partial x}(x,y) \left( \dfrac{z^{\lambda+1}}{\lambda + 1} - \dfrac{1}{2^{\lambda+1}(\lambda + 1)} \right), \\
        \\
       v(x,y,z) = \displaystyle\int_{H/2}^z \dfrac{\partial v}{\partial z} dt = 
        \dfrac{1}{X(\vert\vert p\vert\vert)} \left( \displaystyle{\int}_{H/2}^z \left(2^{\frac{1-\lambda}{2\lambda}}\dfrac{ t\,X(\vert\vert p\vert\vert ) }{f_1(p)}\right)^\lambda dt\right) \dfrac{\partial p}{\partial y}(x,y)
       \\
       \qquad \qquad\qquad \qquad\quad = \dfrac{2^{\frac{1-\lambda}{2}}\left( X(\vert\vert p\vert\vert )\right)^{\lambda-1}}{\left(f_1(p)\right)^\lambda} \dfrac{\partial p}{\partial y}(x,y) \left( \dfrac{z^{\lambda+1}}{\lambda + 1} - \dfrac{1}{2^{\lambda+1}(\lambda + 1)} \right) .  \\        
    \end{cases}
\end{equation}
Moreover, the flow fluxes $q_u$ and $q_v$ are 
\begin{equation}
        q_u(x,y) =  2  \displaystyle\int_0^{1/2} u(x,y,z) dz 
        = - \dfrac{2^{-\frac{1+3\lambda}{2}}}{\lambda + 2} \dfrac{  \left( X(\vert\vert p\vert\vert )\right)^{\lambda-1}}{\left(f_1(p)\right)^\lambda} \dfrac{\partial p}{\partial x}(x,y)   ,      \label{Exp:q_u}
\end{equation}
\begin{equation}
       q_v(x,y) =  2 \displaystyle\int_0^{1/2} v(x,y,z) dz = 
       - \dfrac{2^{-\frac{1+3\lambda}{2}}}{\lambda + 2} \dfrac{  \left( X(\vert\vert p\vert\vert )\right)^{\lambda-1}}{\left(f_1(p)\right)^\lambda} \dfrac{\partial p}{\partial y}(x,y)   . \label{Exp:q_v}
\end{equation}
The flow fluxes in equations \eqref{Exp:q_u} and \eqref{Exp:q_v} must satisfy 
\begin{equation}
    \dfrac{\partial q_u}{\partial x} + \dfrac{\partial q_v}{\partial y} = 0, 
\end{equation}
leading to a close equation for the pressure, namely 
\begin{equation}
     \dfrac{\partial }{\partial x}\left[g(f(p,\vert\vert p\vert\vert ))\dfrac{\partial p}{\partial x}  \right] + \dfrac{\partial }{\partial y}\left[ g(f(p,\vert\vert p\vert\vert )) \dfrac{\partial p }{\partial y}  \right]  = 0 , \label{EqDiff:p_cart}
\end{equation}
where 
\begin{equation}
    g(f(p,\vert\vert p\vert\vert )) = 
    \dfrac{ \left( X(\vert\vert p\vert\vert )\right)^{\lambda-1}}{\left(f_1(p)\right)^\lambda} = \dfrac{\left[\left(   \dfrac{\partial p}{\partial x}\right)^2 + \left(   \dfrac{\partial p}{\partial y}\right)^2\right]^{\frac{\lambda-1}{2}}}{\left(f_1(p)\right)^\lambda}     . \label{Expr:g(X)}
\end{equation}

\section{Particular solution near the corner edge }\label{ps}

In the following we choose the following exponential dependence  
\begin{equation}
    f_1 (r,\theta ) = \exp ( \delta p(r,\theta ) ) , \label{Def:visc-press}
\end{equation}

where $\delta$ is the pressure coefficient ~\cite{Barus1893} and getting inspiration from the work in \cite{Hassager1988-xz} and \cite{Chupin2008},  we proceed by  rewriting equation \eqref{EqDiff:p_cart} in polar coordinates;  one consequently gets:  

\begin{equation}
     \dfrac{\partial }{\partial r}\left[ r  \dfrac{ X^{\lambda-1}(\vert\vert p\vert\vert )}{f_1^\lambda(p)}\dfrac{\partial p}{\partial r} \right] + \dfrac{1}{r}  \dfrac{\partial  }{ \partial \theta } \left(\dfrac{ X^{\lambda-1}(\vert\vert p\vert\vert )}{f_1^\lambda(p)} \dfrac{\partial p}{\partial \theta } \right)  = 0 , \quad     (r,\theta)\in \mathbb{R}_+ \times [ - \alpha , \alpha ] . 
    \label{EqDiff:p_pol}
\end{equation}
Moreover, motivated by the numerical solution presented in \cite{Hassager1988-xz} and the analytical result in \cite{Calusi2024b}, we seek a solution to equation \eqref{EqDiff:p_pol} in the form 
\begin{equation}
    p(r,\theta ) = - \dfrac{1}{\delta}\ln(F(\xi)) , \qquad F(\xi) >0, \label{def:p(r,theta)}
\end{equation}
where $\delta$ is the pressure coefficient ~\cite{Barus1893}, and
\begin{equation}
    \xi(r,\theta) = - \delta r^m \phi(\theta ). 
\end{equation}

The assumption that $F(\xi) >0$ throughout the domain is necessary to ensure the logarithm is well-defined over the domain of interest and simplifies analytical calculations.

\begin{figure}[!h]
    \centering
    \subfloat[ ]{\includegraphics[width=0.47\textwidth]{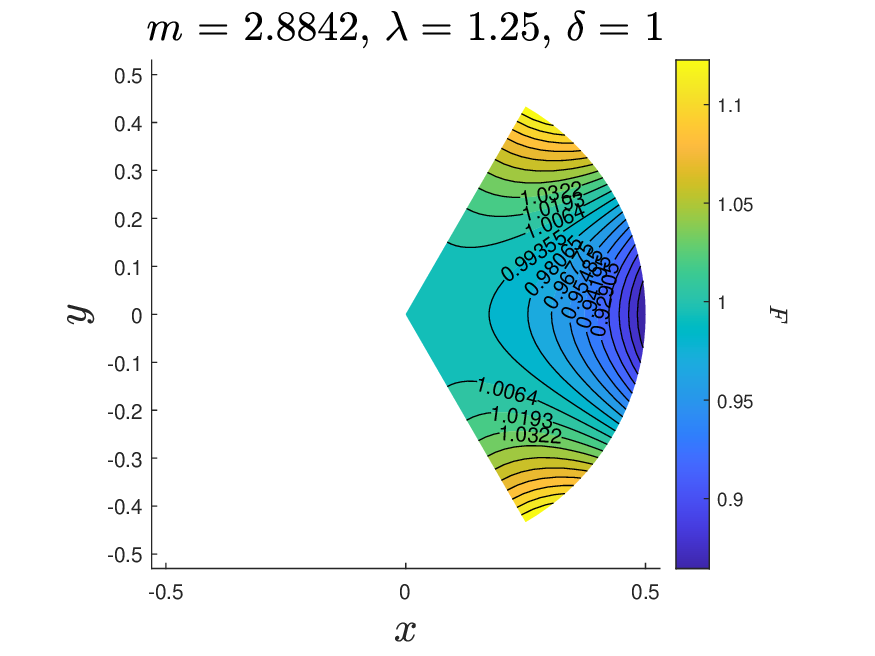} }
    \quad
    \subfloat[ ]{\includegraphics[width=0.47\textwidth]{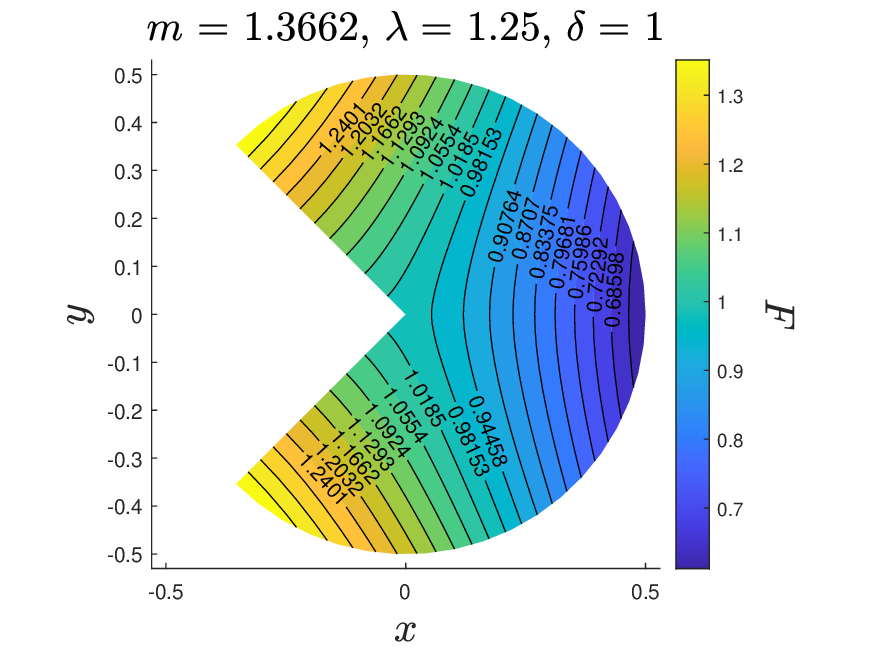}}
     \quad
    \subfloat[ ]{\includegraphics[width=0.47\textwidth]{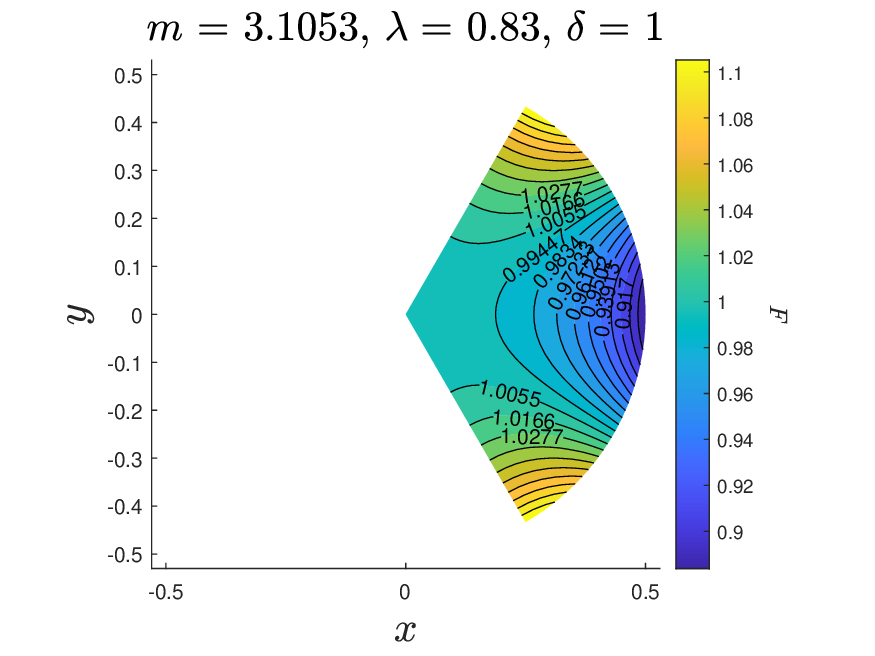}}
     \quad
    \subfloat[ ]{\includegraphics[width=0.47\textwidth]{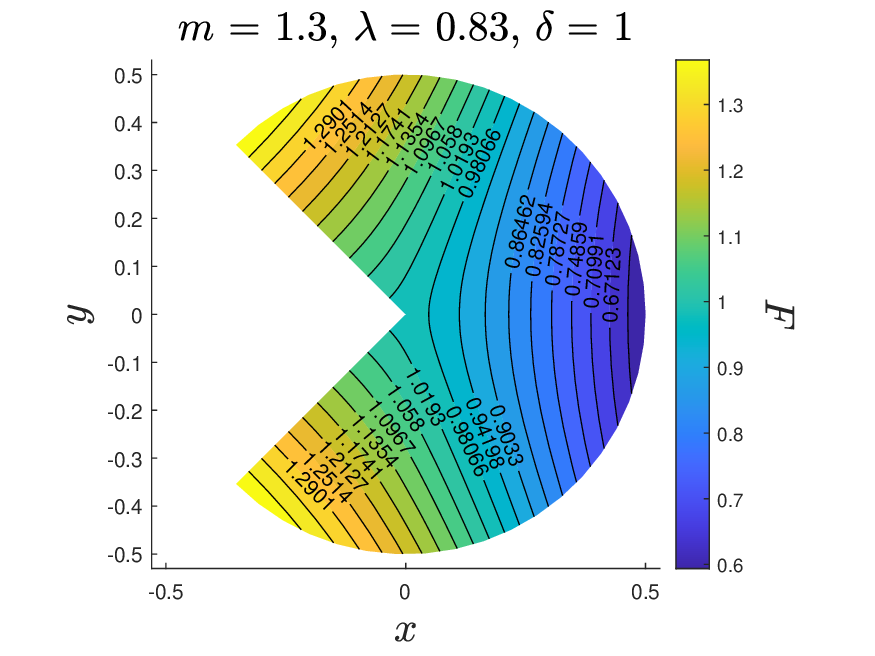}}
    \caption{
Contour plots of the function \( F \) for \( \delta = 1 \) and two different values of \( \lambda \) in the case of symmetric flow : \( \lambda = 1.25 \) (top) and \( \lambda = 0.83 \) (bottom). The angle \( \alpha \) is set to \( \pi/3 \) in the left panel and to \( 3\pi/4 \) in the right panel. In both cases, the function \( F \) remains positive within the domain, confirming the a priori assumption through a posteriori numerical verification.}
    \label{fig:F-SYM-delta-1-lambdas}
\end{figure}

\begin{figure}[!h]
    \centering
    \subfloat[ ]{\includegraphics[width=0.47\textwidth]{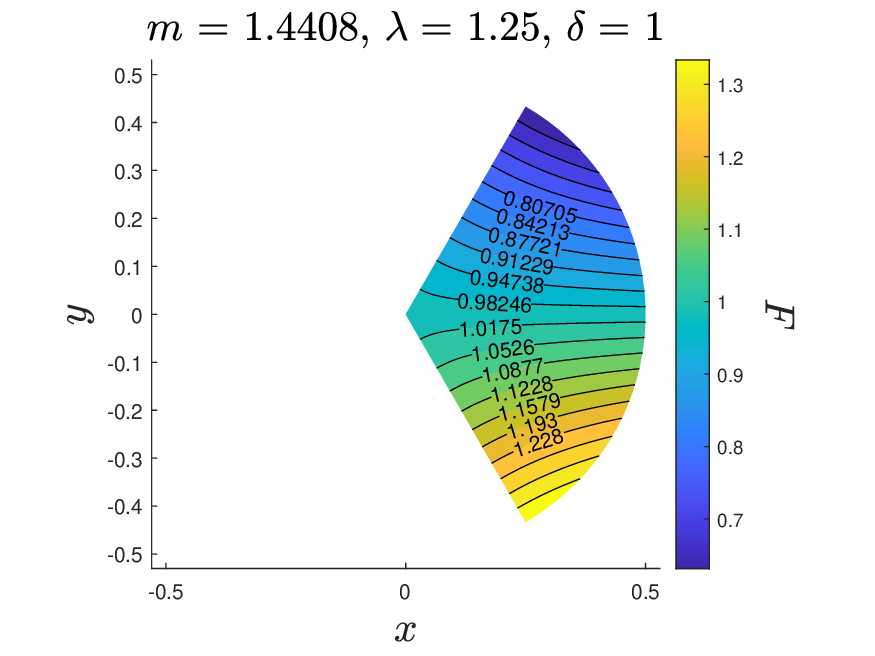} }
    \quad
    \subfloat[ ]{\includegraphics[width=0.47\textwidth]{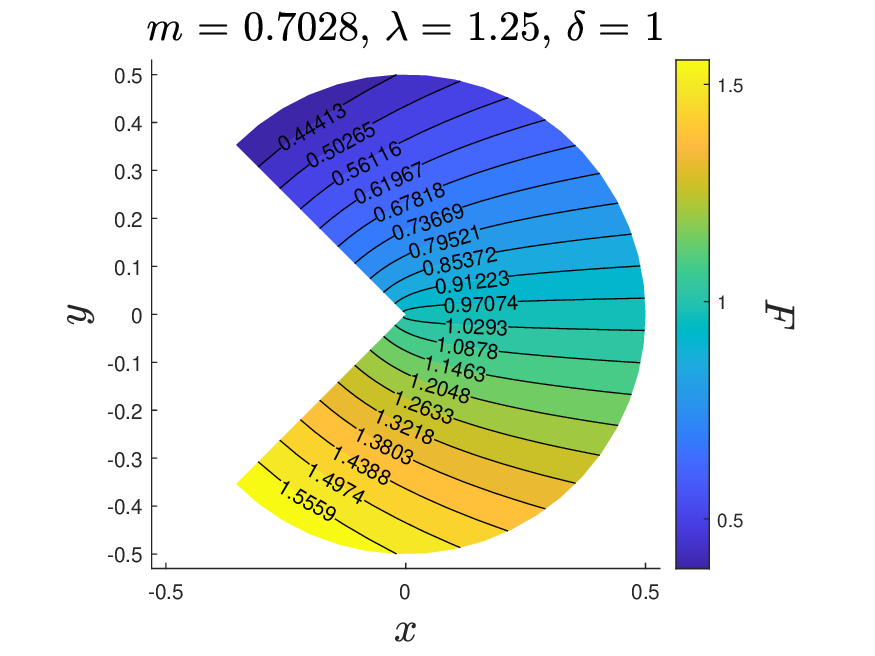}}
     \quad
    \subfloat[ ]{\includegraphics[width=0.47\textwidth]{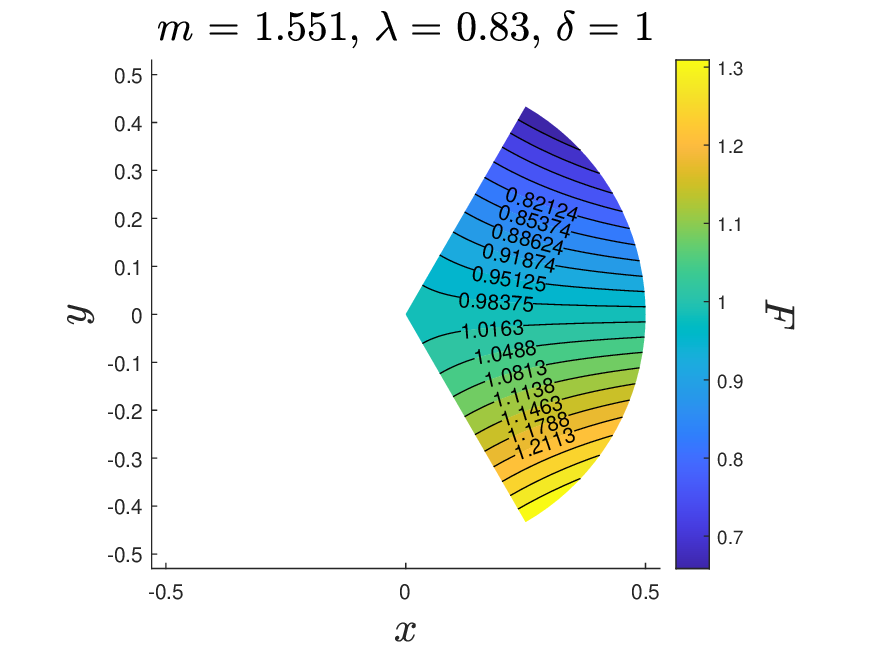}}
     \quad
    \subfloat[ ]{\includegraphics[width=0.47\textwidth]{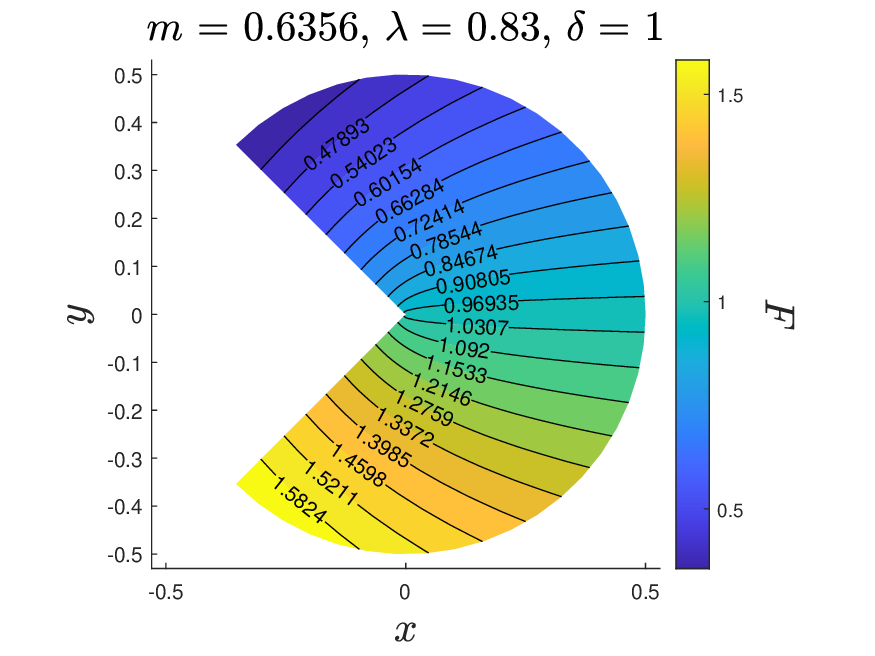}}
    \caption{Contour plots of the function \( F \) for \( \delta = 1 \) and two different values of \( \lambda \) in the case of antisymmetric flow : \( \lambda = 1.25 \) (top) and \( \lambda = 0.83 \) (bottom). The angle \( \alpha \) is set to \( \pi/3 \) in the left panel and to \( 3\pi/4 \) in the right panel. In both cases, the function \( F \) remains positive within the domain, confirming the a priori assumption through a posteriori numerical verification.}
    \label{fig:F-ASYM-delta-1-lambdas}
\end{figure}

The positivity of \( F \) is imposed a priori; numerical simulations consistently show that \( F > 0 \) within the considered parameter range, see Figures \ref{fig:F-SYM-delta-1-lambdas} and  \ref{fig:F-ASYM-delta-1-lambdas}. This a posteriori verification supports the validity of the modeling assumptions under the given conditions. In particular, for the selected parameter values, and taking into account the boundedness of \( \phi \), the function \( F \) remains positive throughout the domain.  

Next, since $X(\vert\vert p\vert \vert)$ can rewritten in polar coordinates as  
\begin{equation}
    X(\vert\vert p\vert\vert) = \sqrt{\left(   \dfrac{\partial p}{\partial r}\right)^2 + \dfrac{1}{r^2} \left(   \dfrac{\partial p}{\partial \theta}\right)^2} = 
    \left[ \left( \dfrac{1}{\delta }\dfrac{\dot{F}}{F} \dfrac{\partial \xi}{\partial r}
     \right)^2 + \dfrac{1}{r^2} \left( \dfrac{1}{\delta }\dfrac{\dot{F}}{ F} \dfrac{\partial \xi}{\partial \theta}
     \right)^2 \right]^{\frac{1}{2}}
    = \left[ r^{2(m-1)} \Dot{F}^2 F^{-2}
     \left( m^2 \phi^2 +  \phi^{\prime ^2}
     \right)\right]^{\frac{1}{2}}  ,  
\end{equation}
equation \eqref{EqDiff:p_pol} rewrites as 
\begin{multline}
     \dfrac{\partial }{\partial r}\Bigg{\lbrace} r^{(m-1)(\lambda - 1) + 1} \dfrac{\left[ 
      \Dot{F}^2 F^{-2}
     \left( m^2 \phi^2 +  \phi^{\prime ^2}
     \right)\right]^{\frac{\lambda-1}{2}}  }{F^{-\lambda  }} \dfrac{\Dot{F}}{F} m r^{m-1} \phi 
          \Bigg\rbrace \\
     + \dfrac{1}{r}  \dfrac{\partial  }{ \partial \theta }\Bigg{\lbrace} r^{(m-1)(\lambda - 1) } \dfrac{\left[ 
      \Dot{F}^2 F^{-2}
     \left( m^2 \phi^2 +  \phi^{\prime ^2}
     \right)\right]^{\frac{\lambda-1}{2}}  }{F^{-\lambda  }} \dfrac{\Dot{F}}{F} r^{m} \phi^\prime 
          \Bigg\rbrace \\
         = \dfrac{\partial }{\partial r}\Bigg{\lbrace} r^{(m-1)\lambda  + 1} 
      \left(\Dot{F}^2 \right)^{\frac{\lambda}{2}}
     \left( m^2 \phi^2 +  \phi^{\prime ^2}
     \right)^{\frac{\lambda-1}{2}}    m  \phi 
          \Bigg\rbrace \\
     + \dfrac{1}{r}  \dfrac{\partial  }{ \partial \theta }\Bigg{\lbrace} r^{(m-1)(\lambda - 1) + m } \left(\Dot{F}^2 \right)^{\frac{\lambda}{2}} \left( m^2 \phi^2 +  \phi^{\prime ^2}
     \right)^{\frac{\lambda-1}{2}} \phi^\prime 
          \Bigg\rbrace = 0 , 
          \label{Eq:diff-F-phi-r}
\end{multline}
where $\Dot{(\,\,\, )} $ and $(\,\,\, )' $ stand for derivative with respect to $\xi$ and to $\theta$, respectively.

Now, we seek a function \( F(\xi) \) such that
\begin{equation}
\left( \dot{F}(\xi)^2 \right)^{\frac{\lambda}{2}} = C_1 = \text{constant}, 
\qquad \Rightarrow \qquad 
F(\xi) = C_1^{-\lambda} \xi + C_2, \label{exp:F}
\end{equation}
where \( C_2 \) is a constant of integration. 
The constants are chosen as \( C_1 = C_2 = 1 \) in order to recover the pressure field \( p \) in the case \( \lambda = 1 \), which corresponds to a piezo-viscous fluid with linear behavior, as discussed in~\cite{Calusi2024b}, namely 
\begin{itemize}
    \item for antisymmetric flow,
    \begin{equation}
        p(r,\theta) = - \dfrac{\ln \left| 1 - \delta r^m \sin (m\theta) \right|}{\delta}, 
    \end{equation}
    \item for symmetric flow,
    \begin{equation}
        p(r,\theta) = - \dfrac{\ln \left| 1 - \delta r^m \cos (m\theta) \right|}{\delta}.
    \end{equation}
\end{itemize}
Unlike our proposed ansatz, which guarantees the positivity of the function inside the logarithm by construction, the latter expressions explicitly include an absolute value to ensure that the solution remains well-defined even when the argument becomes negative.   
Therefore, equation \eqref{Eq:diff-F-phi-r} reduces to 
\begin{multline}
      \ \Bigg\lbrace \dfrac{\partial }{\partial r}\left[ r^{(m-1)\lambda + 1} 
      \left( m^2 \phi^2 +  \phi^{\prime ^2}
     \right)^{\frac{\lambda-1}{2}}    m  \phi 
           \right]
     + \dfrac{1}{r}  \dfrac{\partial  }{ \partial \theta }\left[ r^{(m-1)(\lambda - 1) + m }  \left( m^2 \phi^2 +  \phi^{\prime ^2}
     \right)^{\frac{\lambda-1}{2}} \phi^\prime 
          \right] \Bigg\rbrace \\
         =  r^{(m-1)\lambda }  \Bigg\lbrace 
         m \left[  (m-1)\lambda +1 \right]
      \left( m^2 \phi^2 +  \phi^{\prime ^2}
     \right)^{\frac{\lambda-1}{2}}    \phi 
     + \dfrac{d  }{ d \theta }\left[  \left( m^2 \phi^2 +  \phi^{\prime ^2}
     \right)^{\frac{\lambda-1}{2}} \phi^\prime 
          \right] \Bigg\rbrace 
                    = 0 , 
          \label{Eq:diff-F-phi-r-2}
\end{multline}
i.e.,
\begin{equation}
    m \left[  (m-1)\lambda +1 \right]
      \left( m^2 \phi^2 +  \phi^{\prime ^2}
     \right)^{\frac{\lambda-1}{2}}    \phi \
     + \dfrac{d  }{ d \theta }\left[  \left( m^2 \phi^2 +  \phi^{\prime ^2}
     \right)^{\frac{\lambda-1}{2}} \phi^\prime 
          \right] 
                    = 0 .  \label{EqDiff:p_pol-4}
\end{equation}

From equation \eqref{EqDiff:p_pol-4}, by denoting $1/\lambda = n$, we recover equation (27) of \cite{Hassager1988-xz}, namely 
    \begin{equation}
        \dfrac{d  }{ d \theta } \left[ \left( m^2 \phi^2 +  \phi^{\prime ^2}\right)^{\frac{1-n}{2n}}  \phi ^\prime  \right] + \dfrac{m( m  + n  - 1 ) }{n}\phi \left( m^2 \phi^2 +  \phi^{\prime ^2}
     \right)^{\frac{1-n}{2n}} 
          = 0  .  
    \end{equation}

\noindent Now, we couple equation \eqref{EqDiff:p_pol-4} with the following boundary conditions 
for antisymmetric flows 
          \begin{equation}
              \phi (0) = 0, \quad \mathrm{and} \quad \phi^\prime (\alpha) = 0 , \label{BC:antisym}
          \end{equation}
and for symmetric flows 
          \begin{equation}
              \phi^\prime(0) = 0, \quad \mathrm{and} \quad \phi^\prime (\alpha) = 0 ,\label{BC:sym}
          \end{equation}
and, as in \cite{Hassager1988-xz}, the problem is to find the smallest eigenvalue that satisfies the condition 
          \begin{equation}
          m > m_c =\dfrac{1 - \dfrac{1}{\lambda}}{\dfrac{1}{\lambda} + 1}  , 
          \label{COND:m}
          \end{equation}

where $\dfrac{1}{\lambda}$ of our work corresponds to the power-law index $n$ in \cite{Hassager1988-xz}.  Now, the meaning of this kind of inequalities was first explained in~\cite{has1989}: they were obtained to ensure the integral corresponding to a variational formulation is well defined.  An analysis similar in nature was equally presented in~\cite{Hassager1988-xz} where focus was on power law fluids.  We here capitalize on the aforementioned results.

\begin{figure}[!h]
    \centering
    \subfloat[ ]{\includegraphics[width=0.47\textwidth]{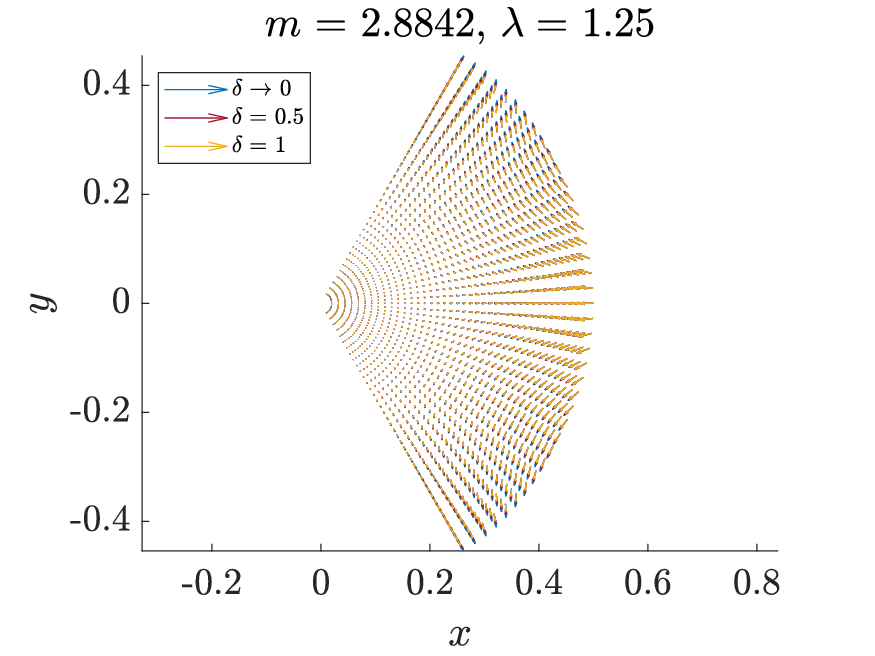} }
    \quad
    \subfloat[ ]{\includegraphics[width=0.47\textwidth]{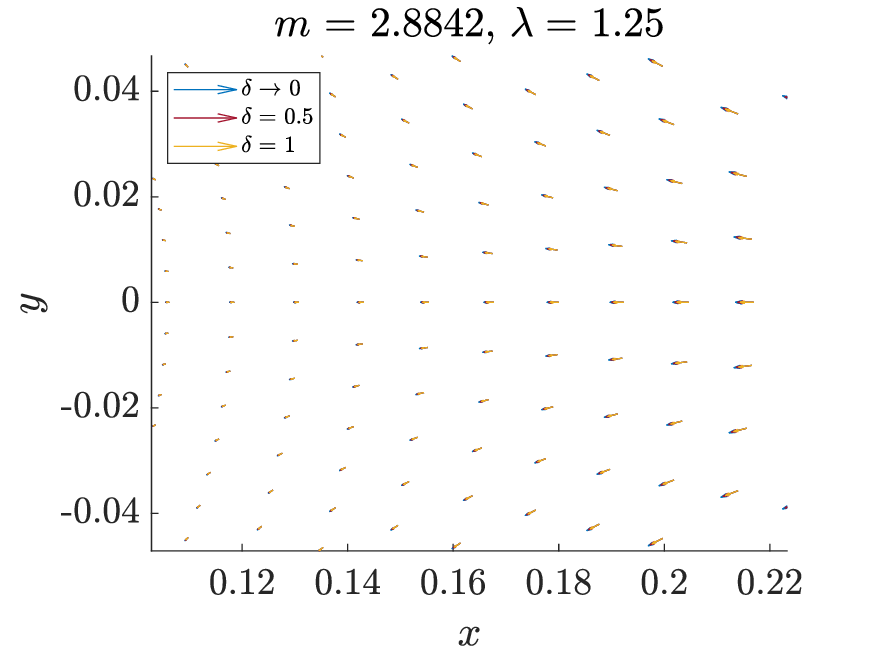}}
     \quad
    \subfloat[ ]{\includegraphics[width=0.47\textwidth]{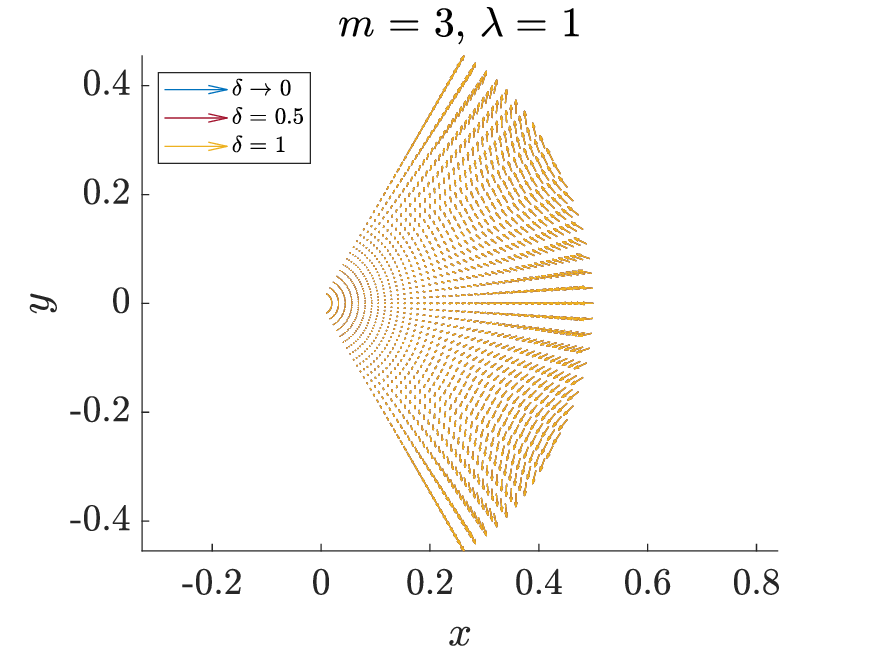}}
     \quad
    \subfloat[ ]{\includegraphics[width=0.47\textwidth]{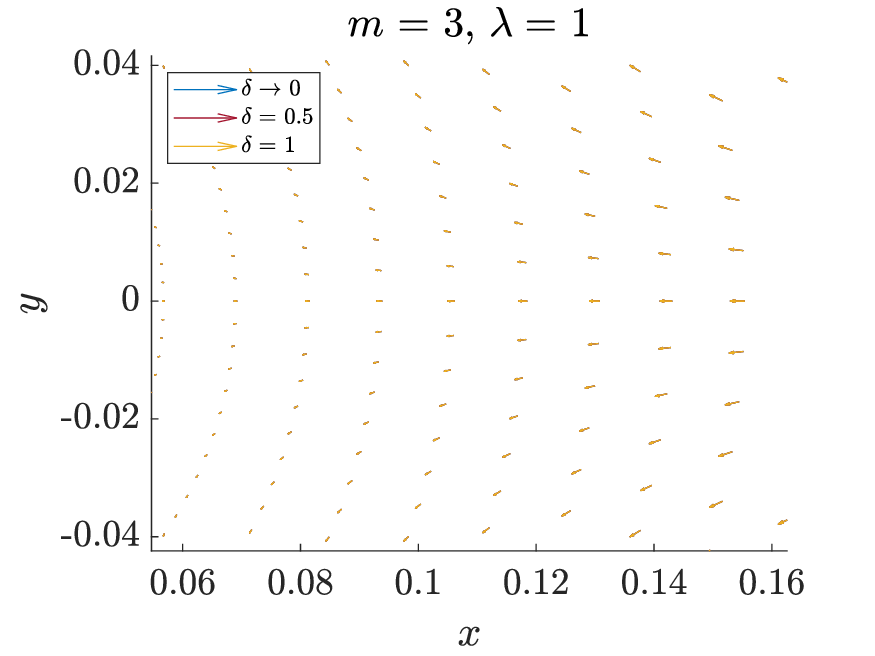}}
    \quad
    \subfloat[ ]{\includegraphics[width=0.47\textwidth]{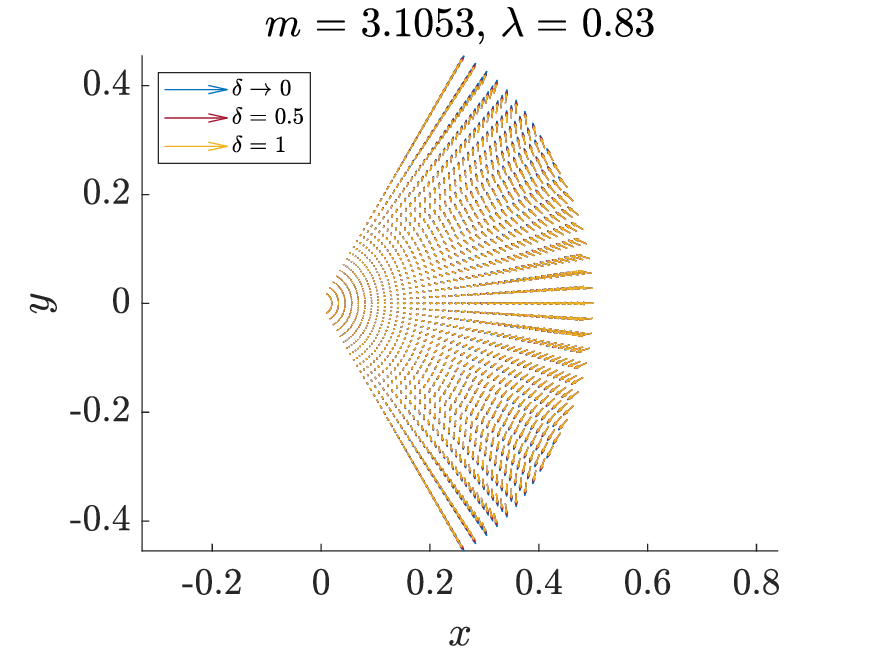}}
     \quad
    \subfloat[ ]{\includegraphics[width=0.47\textwidth]{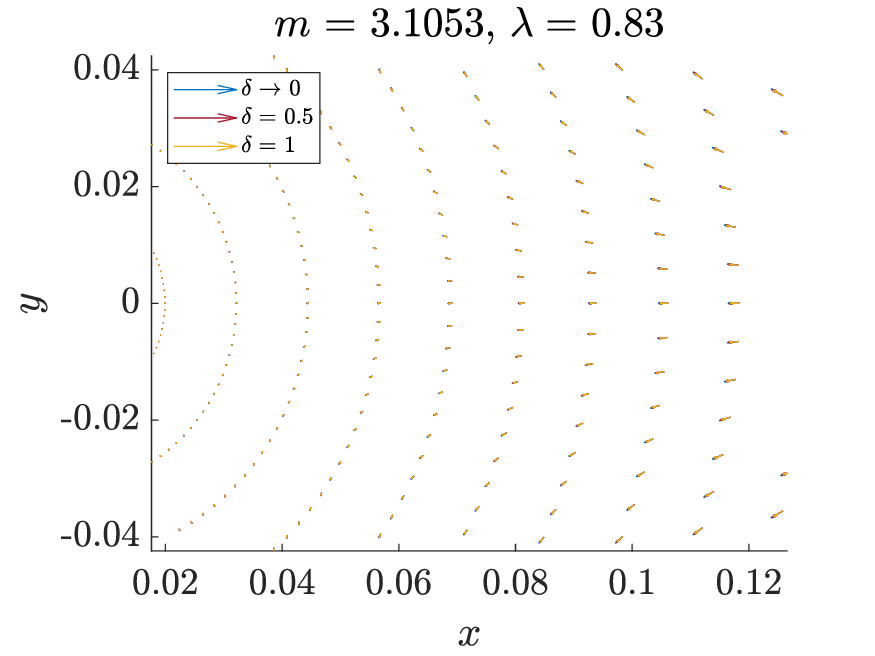}}
    \caption{Comparison of flow components in Cartesian coordinates for a fluid with viscosity depending on both pressure (as defined in equation~\eqref{Def:visc-press}) and shear rate. Results are shown for three values of the nonlinearity exponent $\lambda$, and for three levels of pressure dependence: $\delta \to 0$ (no piezo-viscous effect), $\delta = 0.5$, and $\delta = 1$. Each plot displays the superposition of the three corresponding $\delta$ values. The left column refers to symmetric flow in a non-reentrant angular geometry with opening angle $\alpha = \frac{\pi}{3}$, while the right column provides zoomed-in views near the corner region. The classical Newtonian case corresponds to $\lambda = 1$ and $\delta \to 0$. }\label{fig:SYM_pi3}
\end{figure}

\begin{figure}[!h]
    \centering
    \subfloat[ ]{\includegraphics[width=0.47\textwidth]{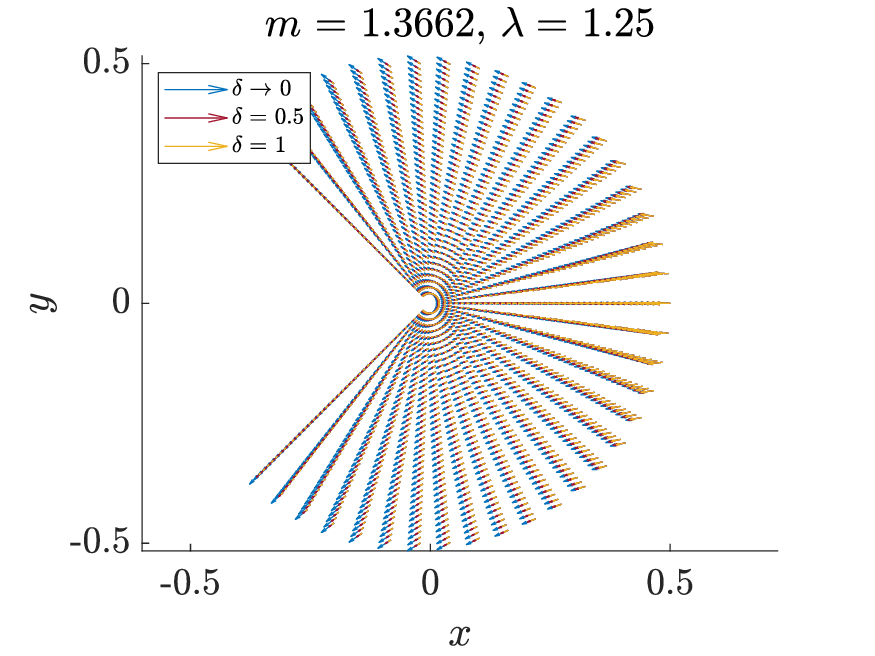} }
    \quad
    \subfloat[ ]{\includegraphics[width=0.35\textwidth]{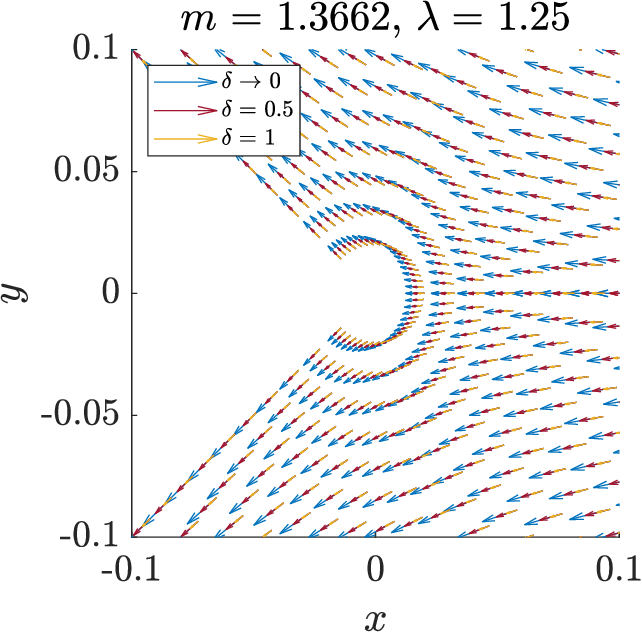}}
     \quad
    \subfloat[ ]{\includegraphics[width=0.47\textwidth]{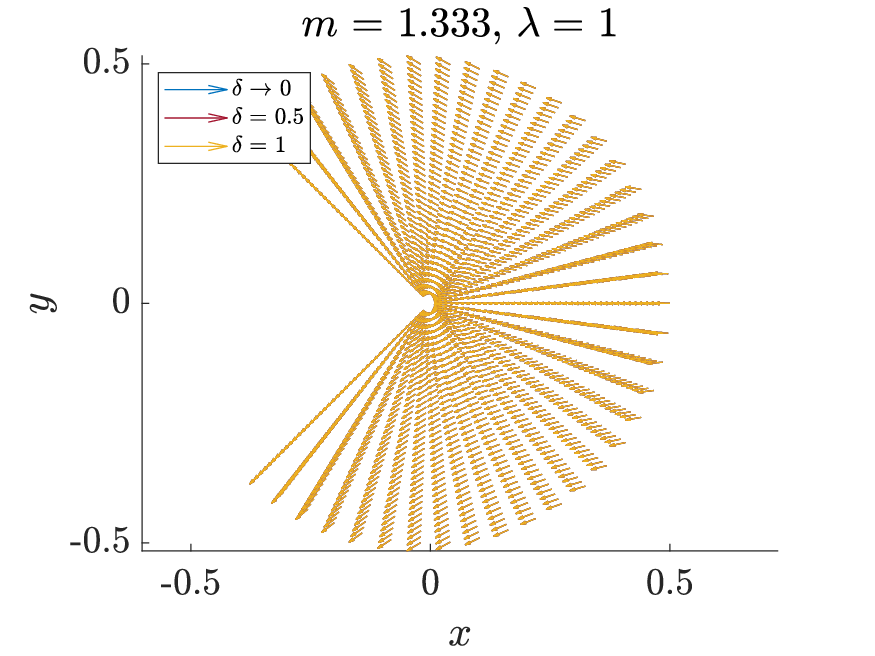}}
     \quad
    \subfloat[ ]{\includegraphics[width=0.35\textwidth]{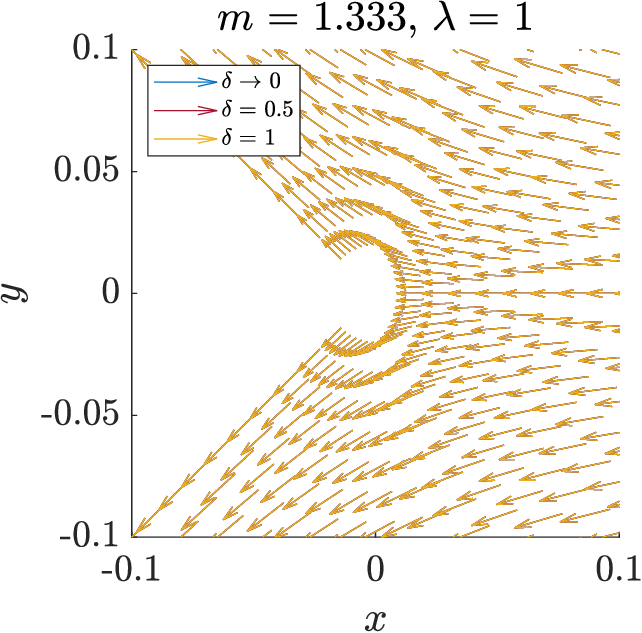}}
    \quad
    \subfloat[ ]{\includegraphics[width=0.47\textwidth]{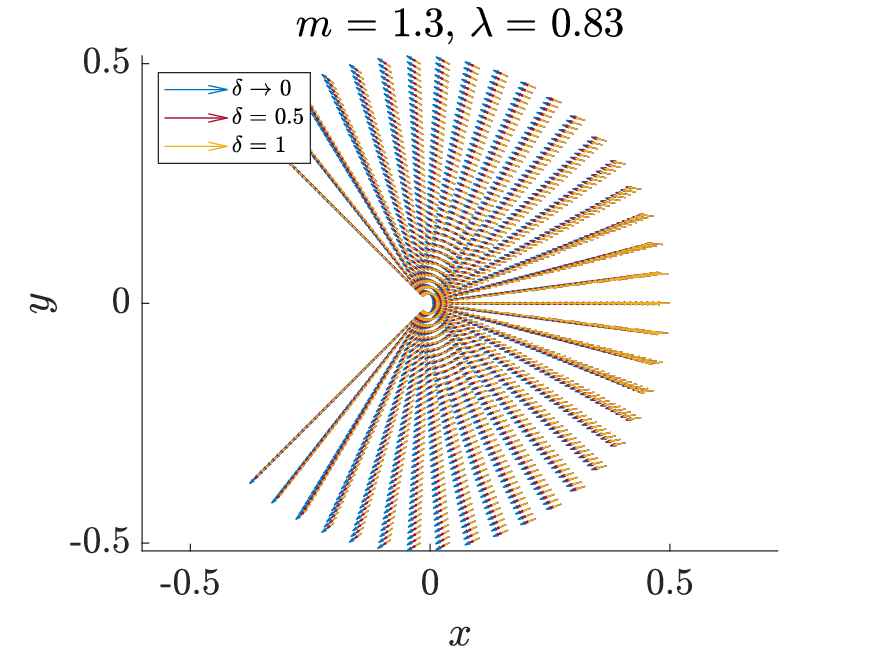}}
     \quad
    \subfloat[ ]{\includegraphics[width=0.35\textwidth]{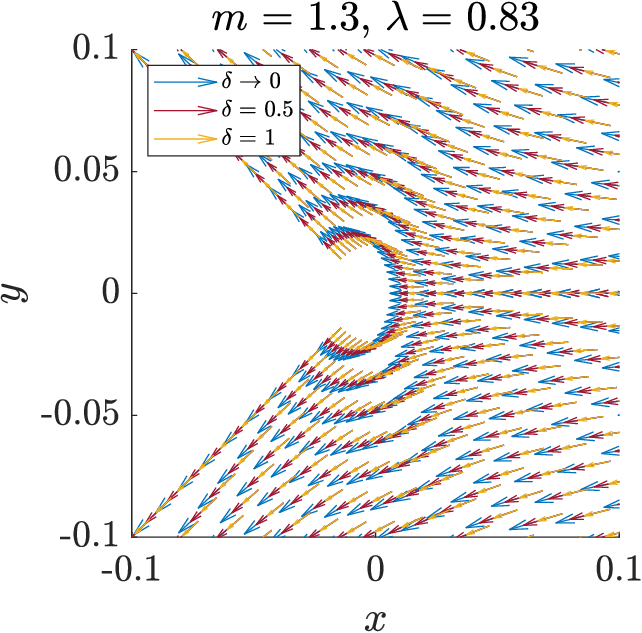}}
    \caption{Comparison of flow components in Cartesian coordinates for a fluid with viscosity depending on both pressure (as defined in equation~\eqref{Def:visc-press}) and shear rate. Results are shown for three values of the nonlinearity exponent $\lambda$, and for three levels of pressure dependence: $\delta \to 0$ (no piezo-viscous effect), $\delta = 0.5$, and $\delta = 1$. Each plot displays the superposition of the three corresponding $\delta$ values. The left column refers to symmetric flow in a non-reentrant angular geometry with opening angle $\alpha = \frac{3\pi}{4}$, while the right column provides zoomed-in views near the corner region. The classical Newtonian case corresponds to $\lambda = 1$ and $\delta \to 0$.}
    \label{fig:fig:SYM_3pi4}
\end{figure}

\begin{figure}[!h]
    \centering
    \subfloat[ ]{\includegraphics[width=0.47\textwidth]{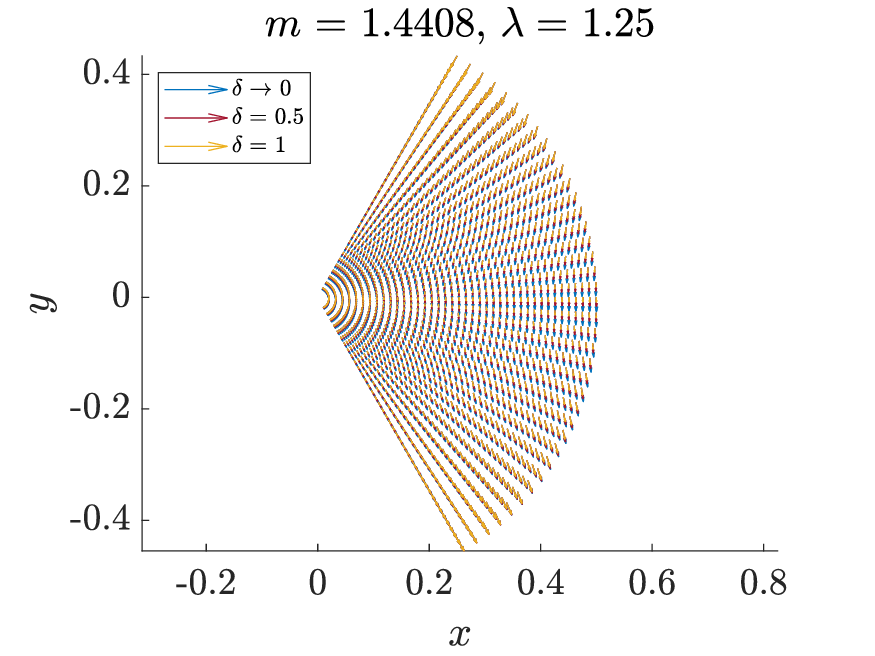} }
    \quad
    \subfloat[ ]{\includegraphics[width=0.35\textwidth]{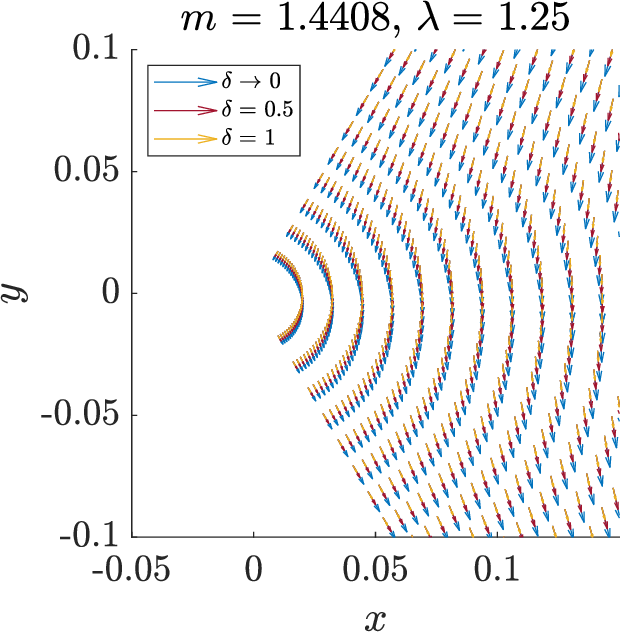}}
     \quad
    \subfloat[ ]{\includegraphics[width=0.47\textwidth]{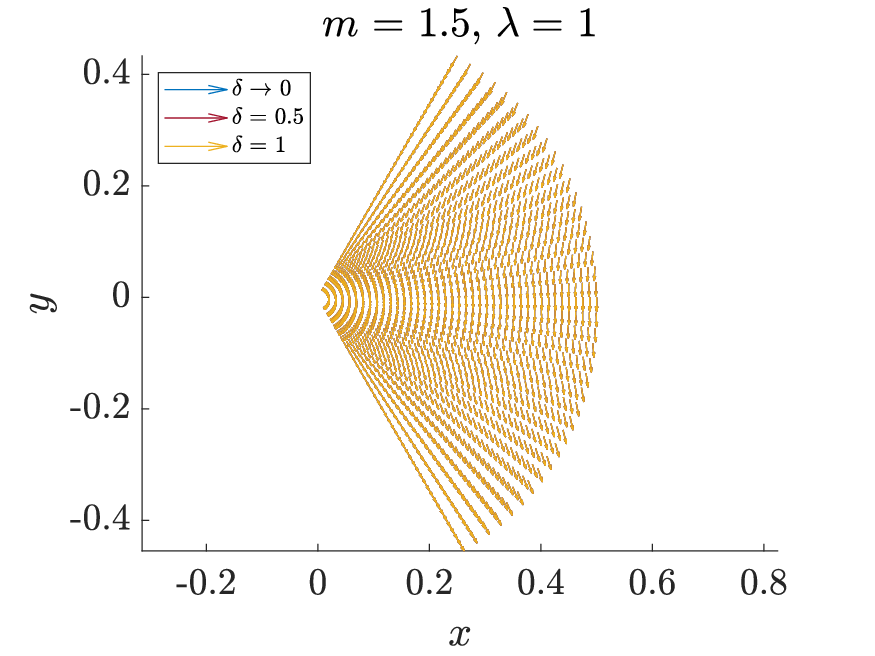}}
     \quad
    \subfloat[ ]{\includegraphics[width=0.35\textwidth]{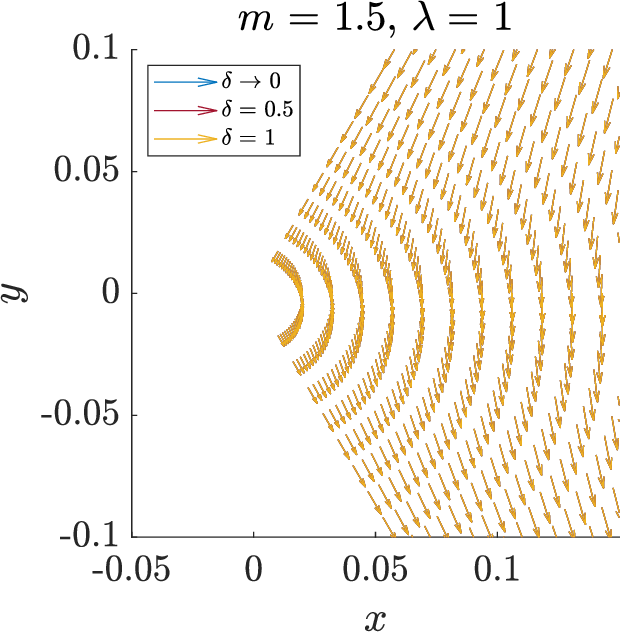}}
    \quad
    \subfloat[ ]{\includegraphics[width=0.47\textwidth]{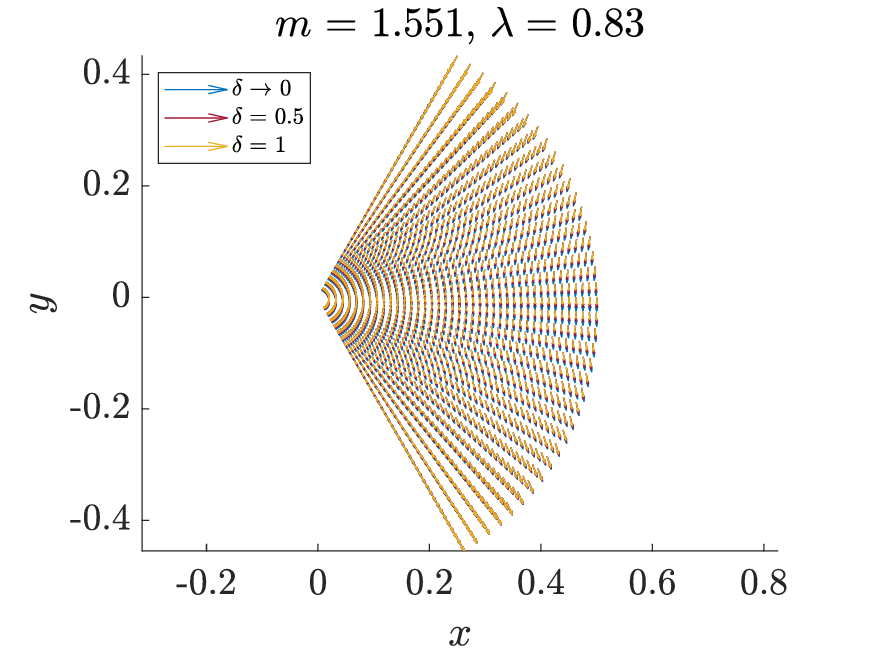}}
     \quad
    \subfloat[ ]{\includegraphics[width=0.35\textwidth]{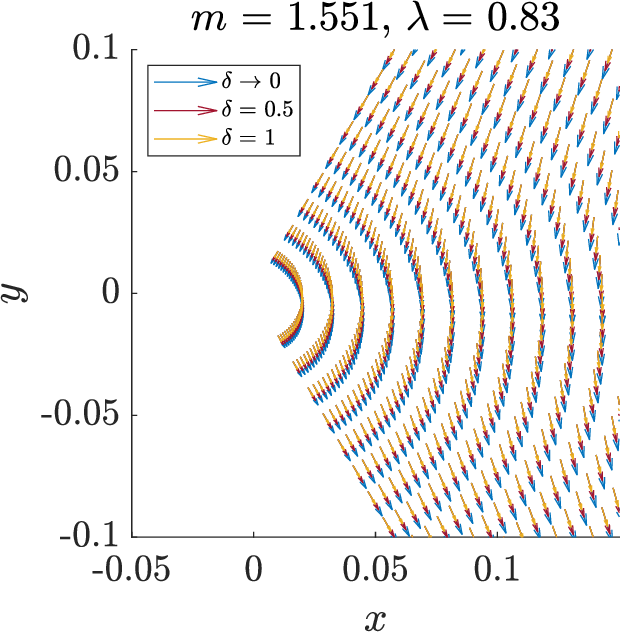}}
    \caption{Comparison of flow components in Cartesian coordinates for a fluid with viscosity depending on both pressure (as defined in equation~\eqref{Def:visc-press}) and shear rate. Results are shown for three values of the nonlinearity exponent $\lambda$, and for three levels of pressure dependence: $\delta \to 0$ (no piezo-viscous effect), $\delta = 0.5$, and $\delta = 1$. Each plot displays the superposition of the three corresponding $\delta$ values. The left column refers to antisymmetric flow in a non-reentrant angular geometry with opening angle $\alpha = \frac{\pi}{3}$, while the right column provides zoomed-in views near the corner region. The classical Newtonian case corresponds to $\lambda = 1$ and $\delta \to 0$.}
    \label{fig:ASYM_pi3}
\end{figure}

\begin{figure}[!h]
    \centering
    \subfloat[ ]{\includegraphics[width=0.47\textwidth]{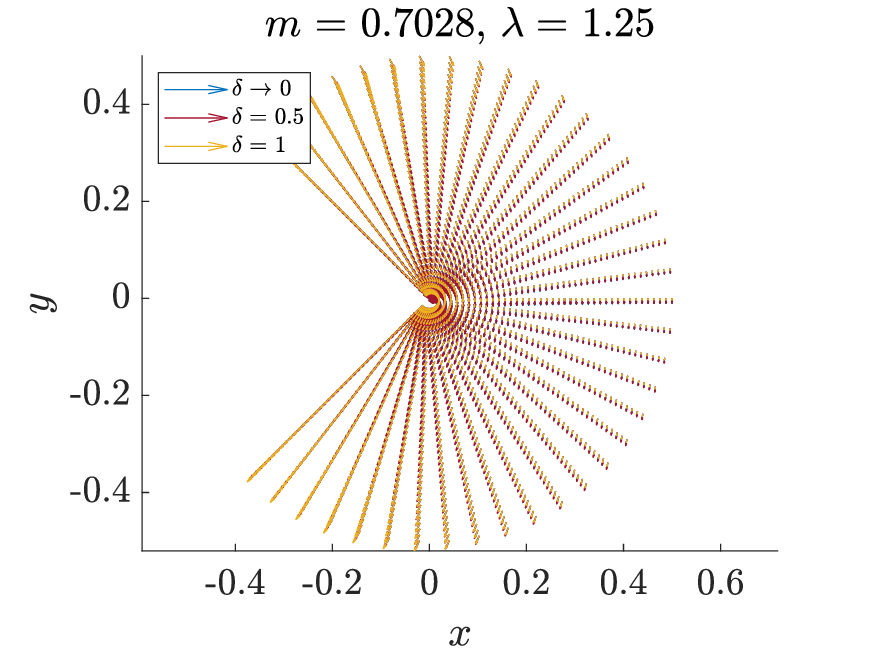} }
    \quad
    \subfloat[ ]{\includegraphics[width=0.35\textwidth]{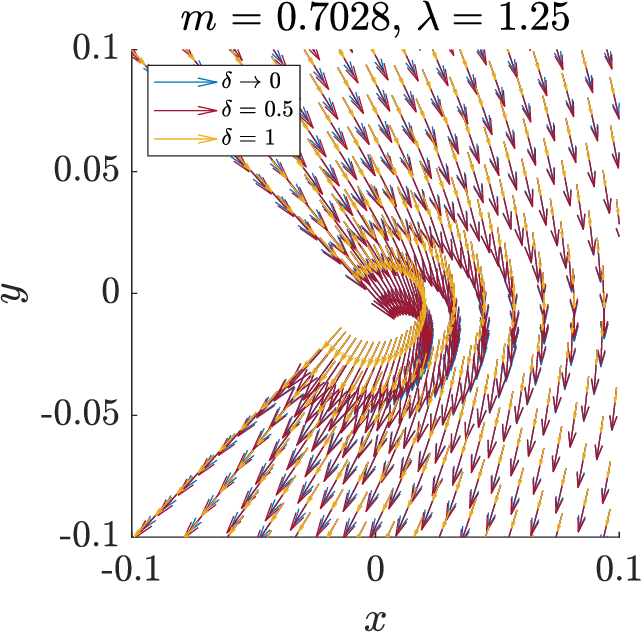}}
     \quad
    \subfloat[ ]{\includegraphics[width=0.47\textwidth]{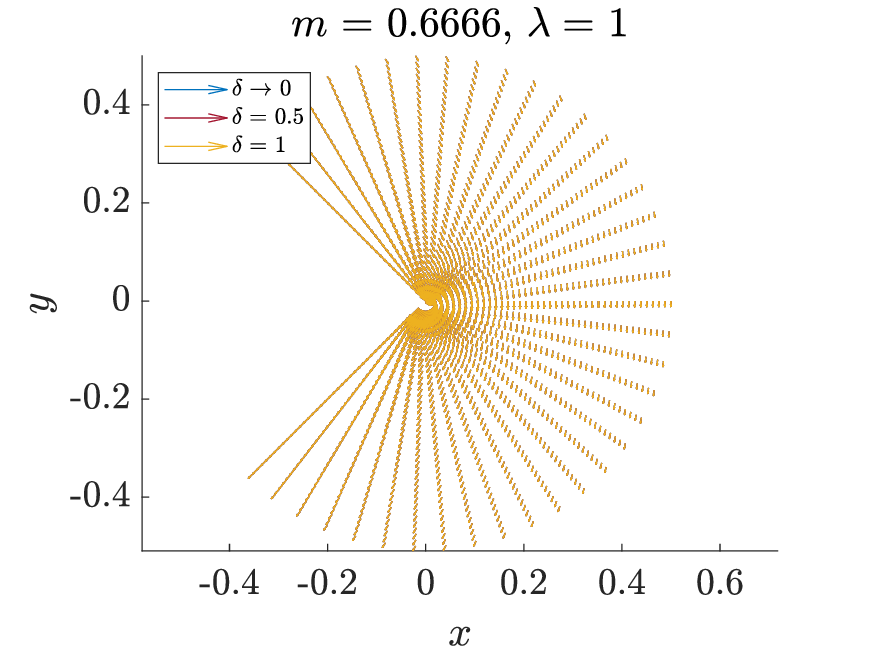}}
     \quad
    \subfloat[ ]{\includegraphics[width=0.35\textwidth]{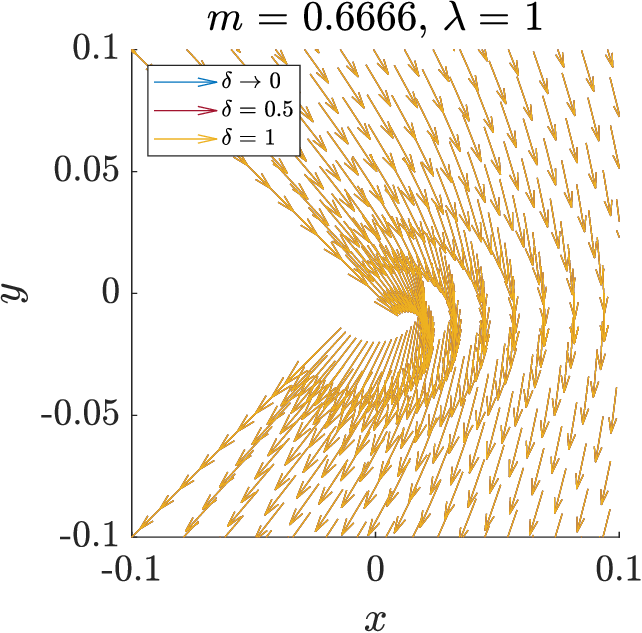}}
    \quad
    \subfloat[ ]{\includegraphics[width=0.47\textwidth]{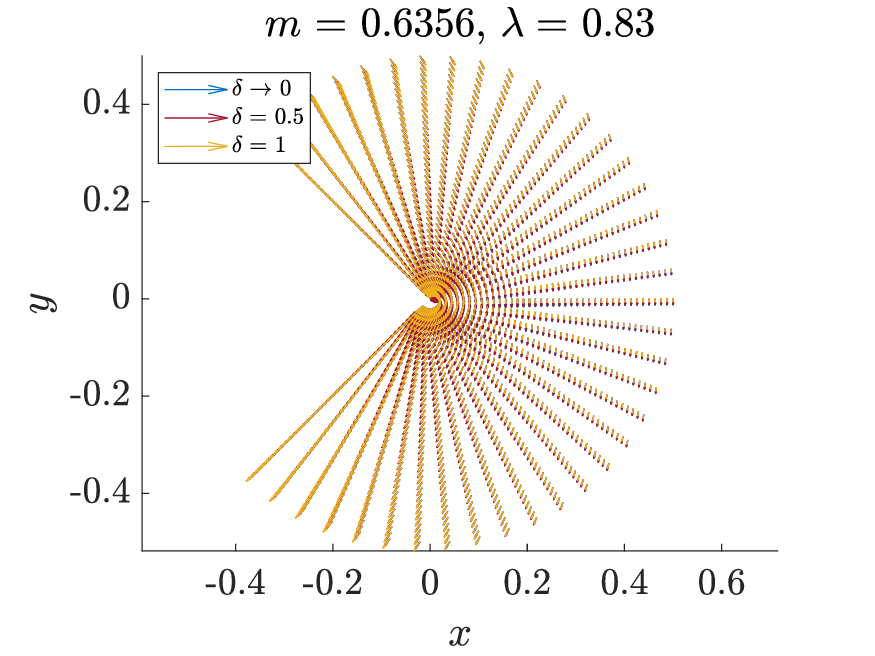}}
     \quad
    \subfloat[ ]{\includegraphics[width=0.35\textwidth]{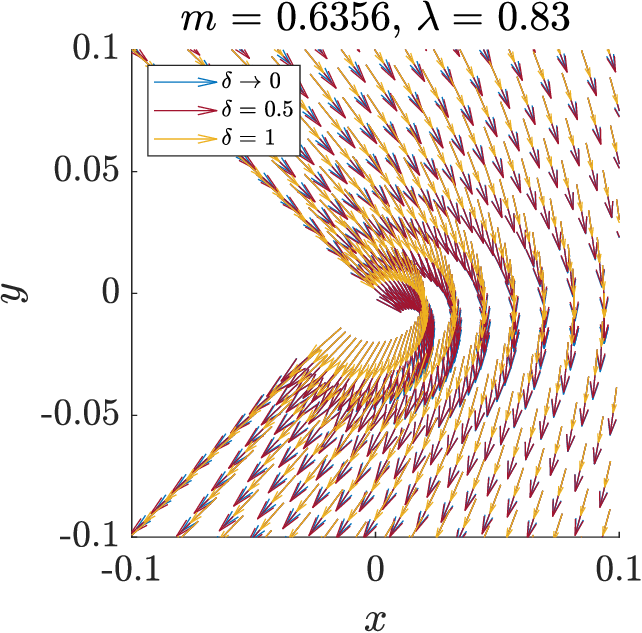}}
    \caption{Comparison of flow components in Cartesian coordinates for a fluid with viscosity depending on both pressure (as defined in equation~\eqref{Def:visc-press}) and shear rate. Results are shown for three values of the nonlinearity exponent $\lambda$, and for three levels of pressure dependence: $\delta \to 0$ (no piezo-viscous effect), $\delta = 0.5$, and $\delta = 1$. Each plot displays the superposition of the three corresponding $\delta$ values. The left column refers to antisymmetric flow in a non-reentrant angular geometry with opening angle $\alpha = \frac{3\pi}{4}$, while the right column provides zoomed-in views near the corner region. The classical Newtonian case corresponds to $\lambda = 1$ and $\delta \to 0$.}
    \label{fig:ASYM_3pi4}
\end{figure}

\begin{figure}[!h]
    \centering
    \subfloat[ ]{\includegraphics[width=0.47\textwidth]{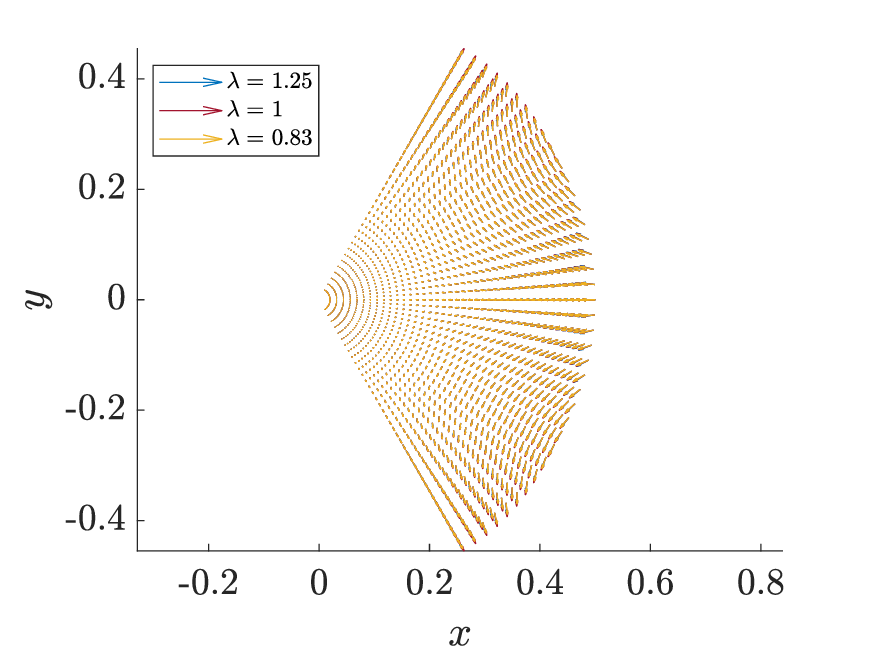} }
    \quad
    \subfloat[ ]{\includegraphics[width=0.44\textwidth]{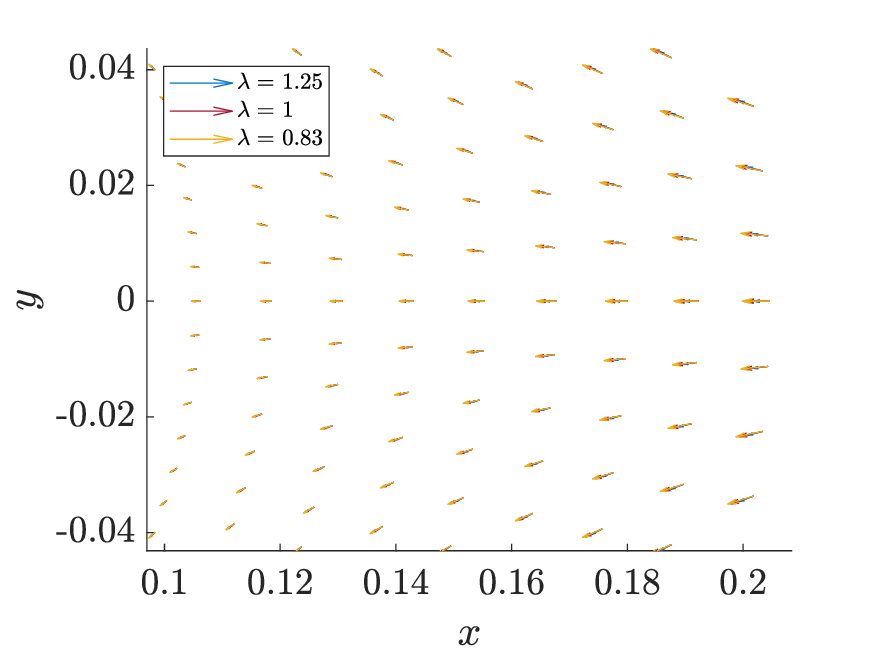}}
     \quad
    \subfloat[ ]{\includegraphics[width=0.47\textwidth]{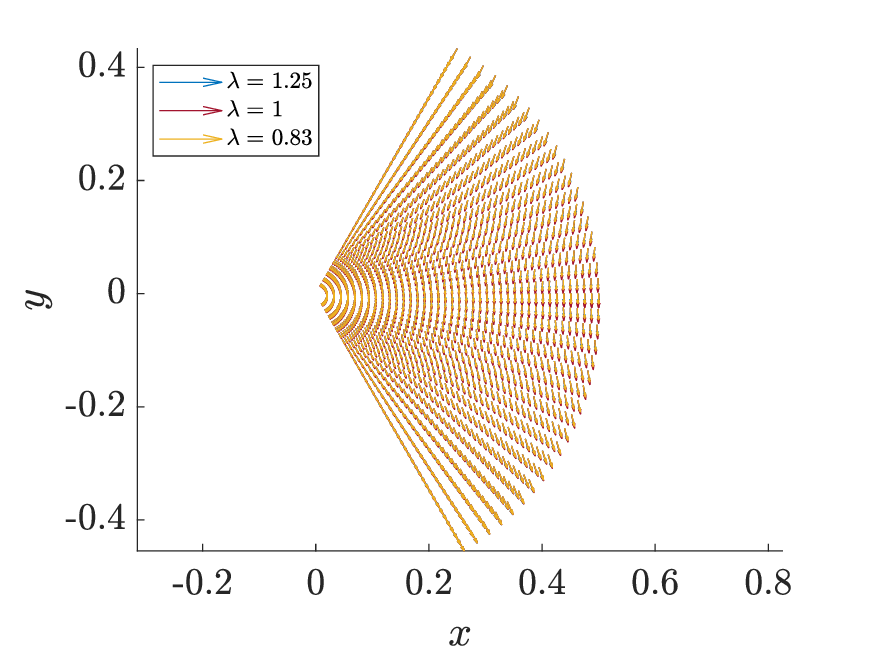}}
     \quad\qquad \qquad
    \subfloat[ ]{\includegraphics[width=0.35\textwidth]{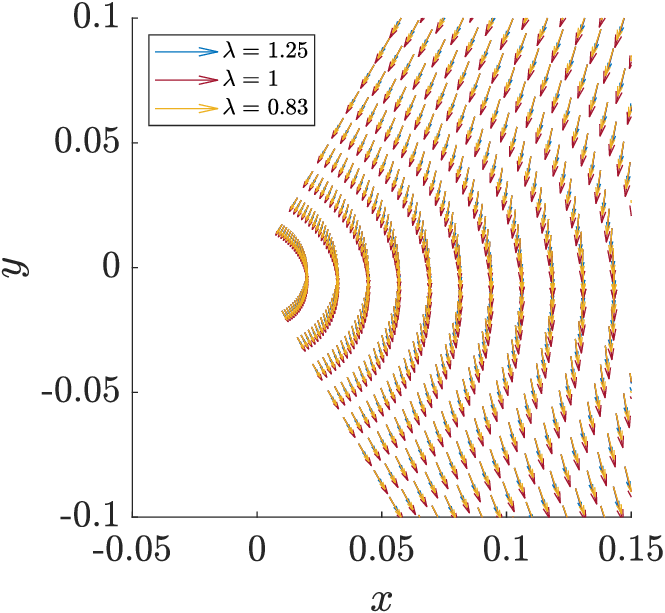}}
    \caption{Comparison of flow fluxes in Cartesian coordinates for a fluid with viscosity dependent on both pressure and shear rate (the pressure dependence is defined in equation~\eqref{Def:visc-press}), for three different values of the nonlinearity exponent $\lambda$. The left column shows results for a non-reentrant angular geometry with opening angle $\alpha = \frac{\pi}{3}$ when $\delta = 0.5$: symmetric flow  (top), and antisymmetric flow (bottom). The right column displays the corresponding zoomed-in views near the corner region.}
    \label{fig:SYM-ASYM-pi3-delta-05}
\end{figure}

\begin{figure}[!h]
    \centering
    \subfloat[ ]{\includegraphics[width=0.47\textwidth]{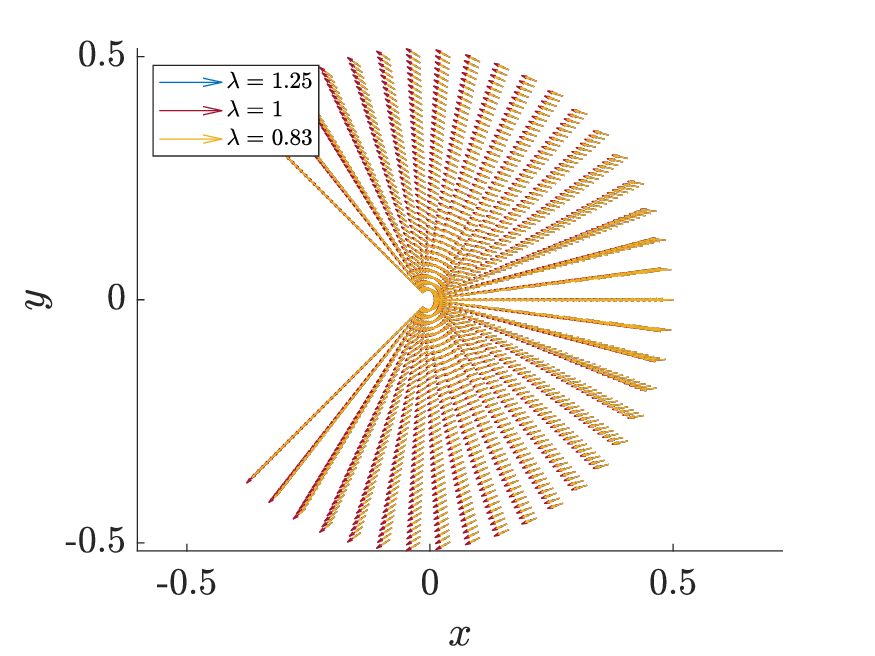} }
    \quad
    \subfloat[ ]{\includegraphics[width=0.35\textwidth]{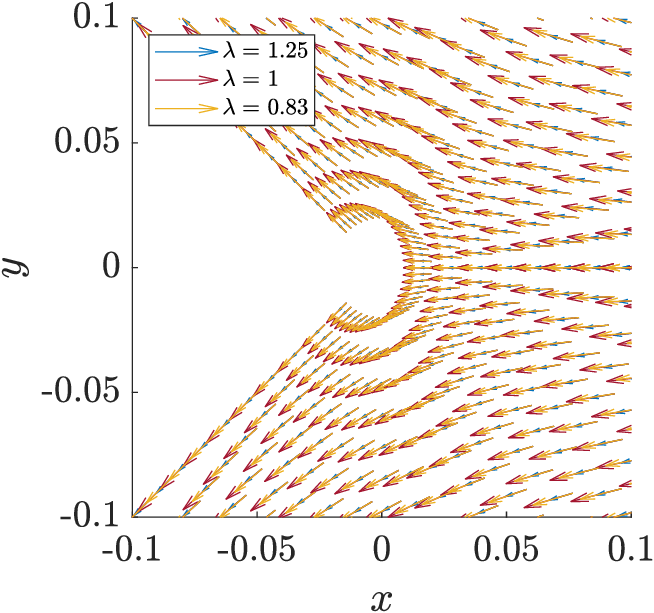}}
     \quad
    \subfloat[ ]{\includegraphics[width=0.47\textwidth]{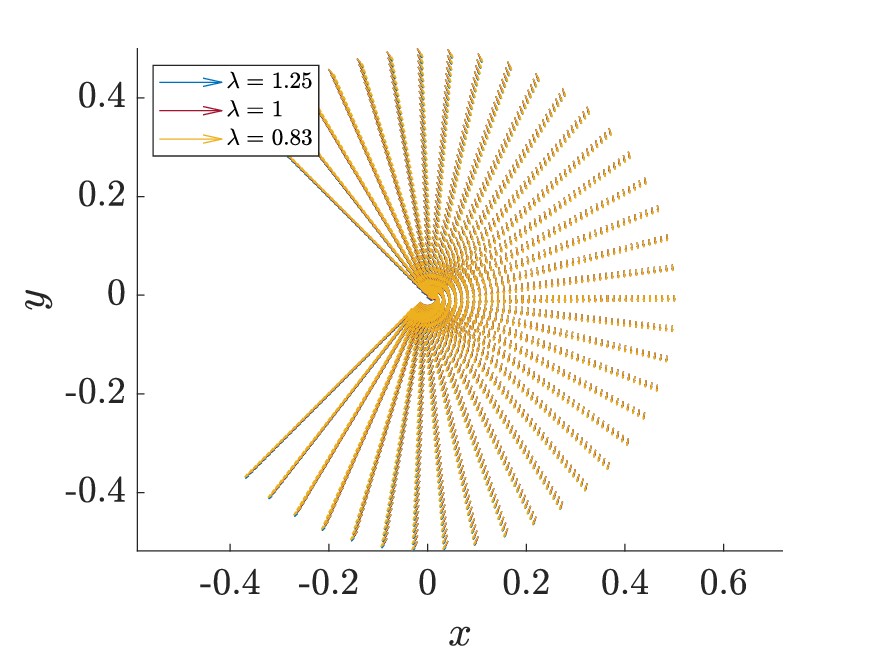}}
     \quad
    \subfloat[ ]{\includegraphics[width=0.35\textwidth]{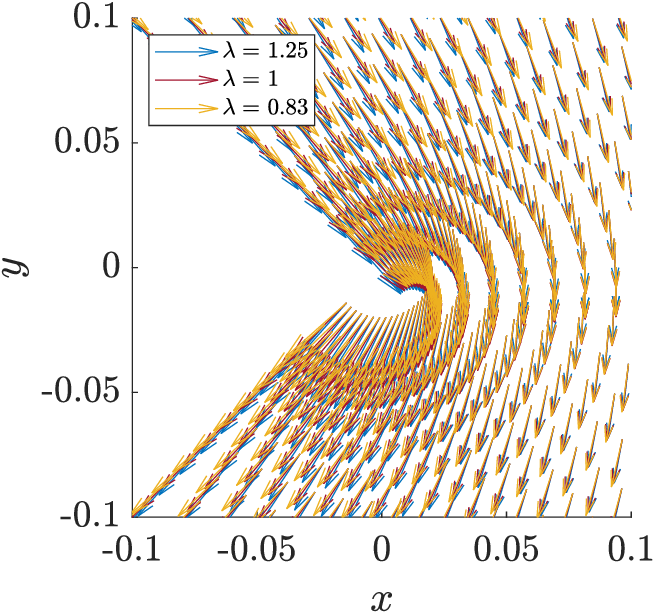}}
    \caption{Comparison of flow fluxes in Cartesian coordinates for a fluid with viscosity dependent on both pressure and shear rate (the pressure dependence is defined in equation~\eqref{Def:visc-press}), for three different values of the nonlinearity exponent $\lambda$. The left column shows results for a non-reentrant angular geometry with opening angle $\alpha = \frac{3\pi}{4}$ when $\delta = 0.5$: symmetric flow  (top), and antisymmetric flow (bottom). The right column displays the corresponding zoomed-in views near the corner region.}
    \label{fig:SYM-ASYM-3pi4-delta-05}
\end{figure}

\begin{figure}[!h]
    \centering
      \subfloat[ ]{\includegraphics[width=0.47\textwidth]{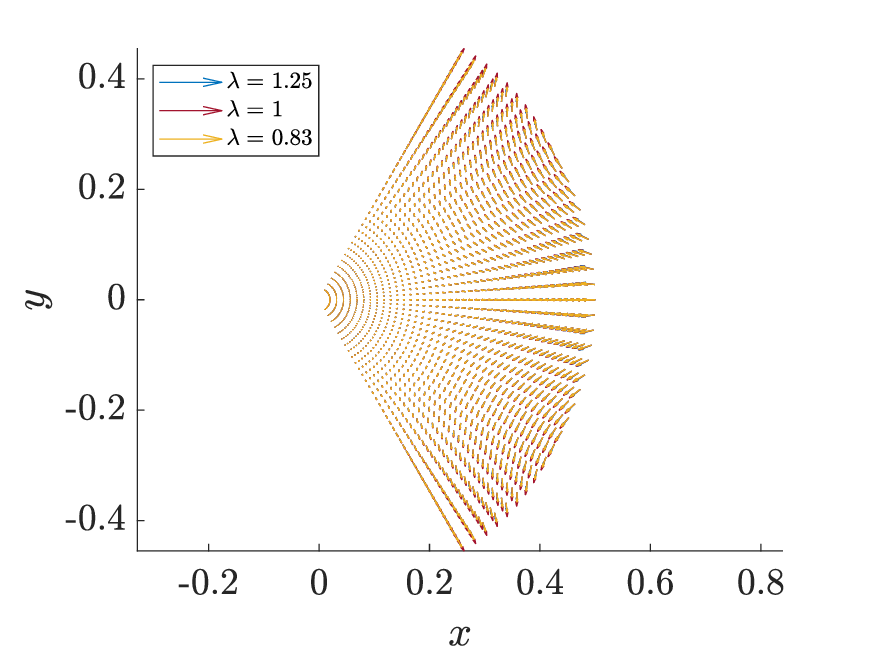} }
    \quad
    \subfloat[ ]{\includegraphics[width=0.44\textwidth]{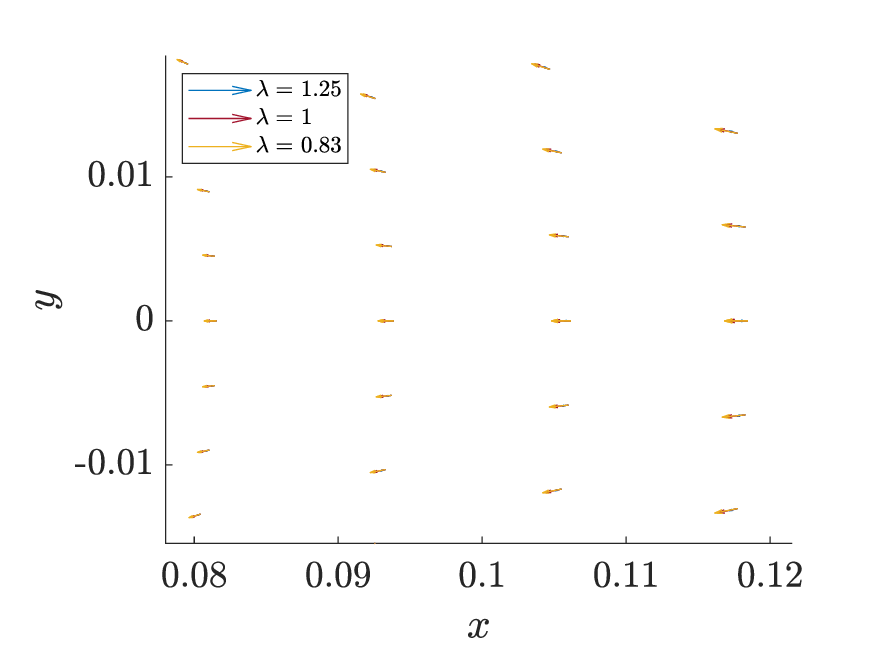}}
     \quad
    \subfloat[ ]{\includegraphics[width=0.47\textwidth]{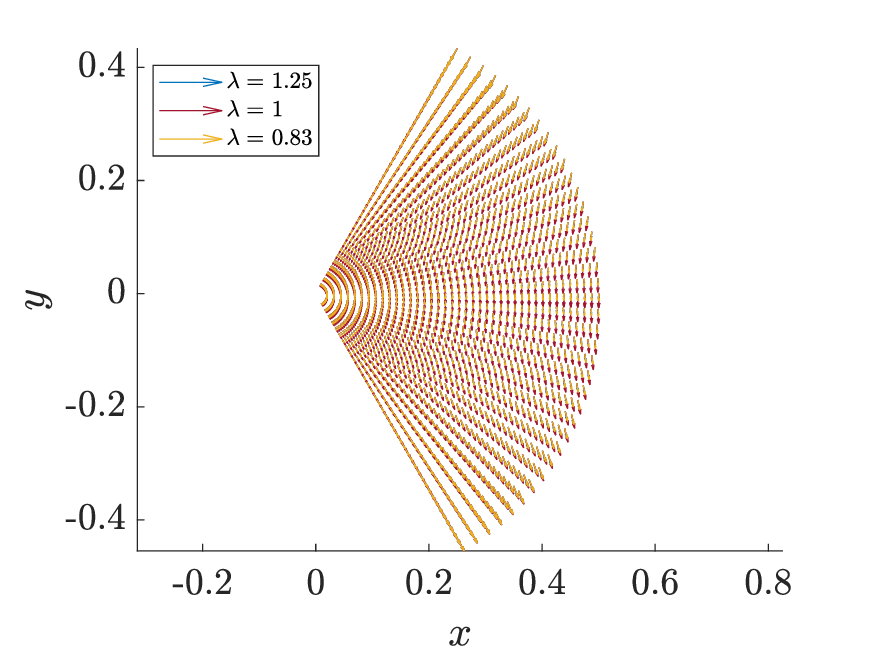}}
     \quad\qquad \qquad
    \subfloat[ ]{\includegraphics[width=0.35\textwidth]{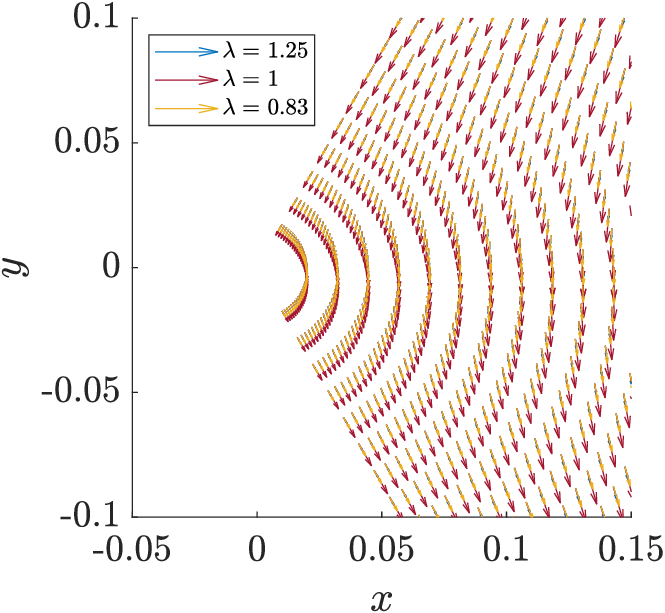}}
    \caption{Comparison of flow fluxes in Cartesian coordinates for a fluid with viscosity dependent on both pressure and shear rate (the pressure dependence is defined in equation~\eqref{Def:visc-press}), for three different values of the nonlinearity exponent $\lambda$. The left column shows results for a non-reentrant angular geometry with opening angle $\alpha = \frac{\pi}{3}$ when $\delta = 1$: symmetric flow  (top), and antisymmetric flow (bottom). The right column displays the corresponding zoomed-in views near the corner region.}
    \label{fig:SYM-ASYM-pi3}
\end{figure}

\begin{figure}[!h]
    \centering
    \subfloat[ ]{\includegraphics[width=0.47\textwidth]{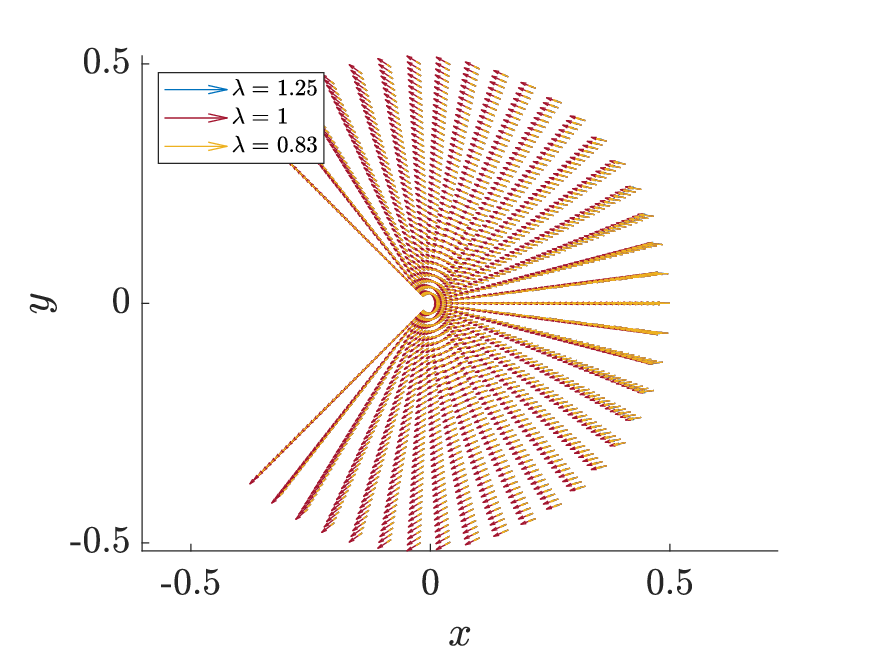} }
    \quad
    \subfloat[ ]{\includegraphics[width=0.35\textwidth]{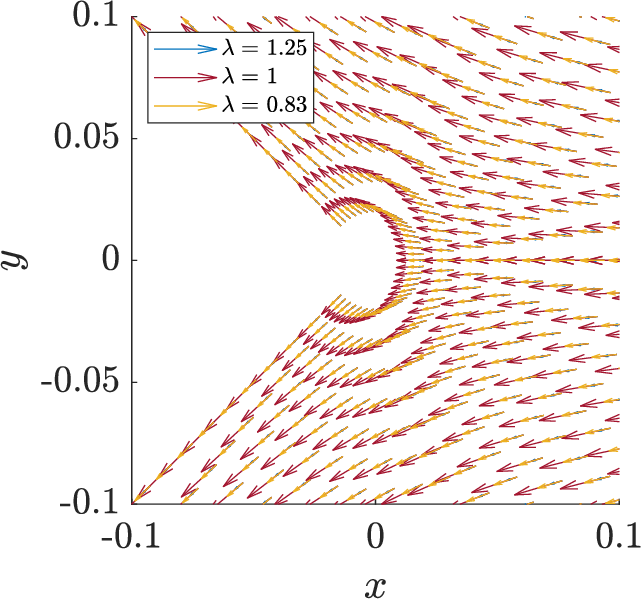}}
     \quad
    \subfloat[ ]{\includegraphics[width=0.47\textwidth]{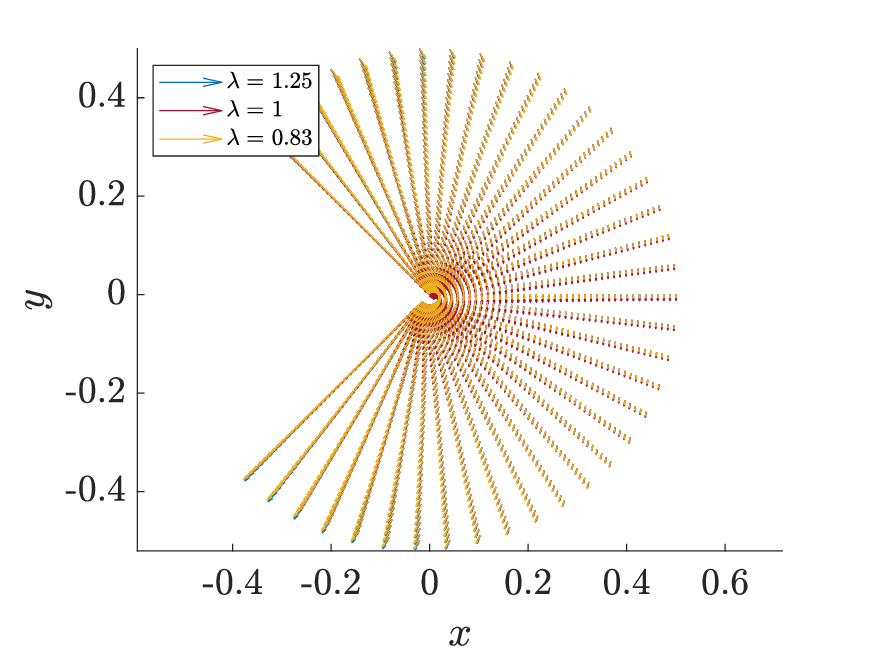}}
     \quad
    \subfloat[ ]{\includegraphics[width=0.35\textwidth]{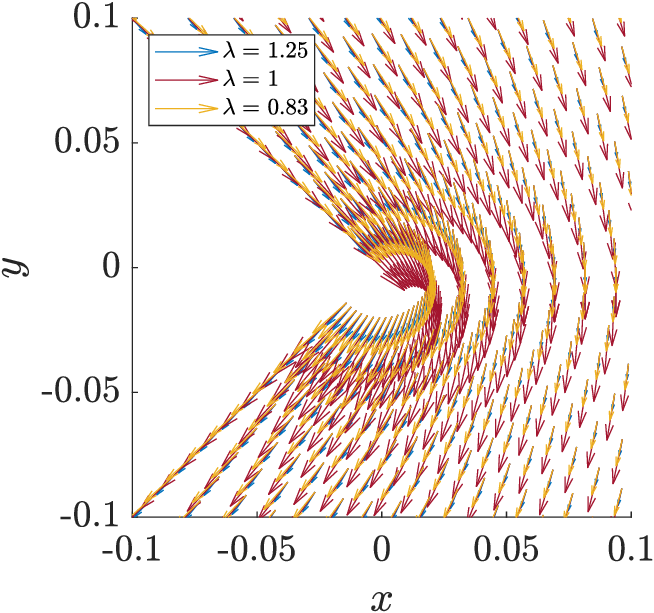}}
    \caption{Comparison of flow fluxes in Cartesian coordinates for a fluid with viscosity dependent on both pressure and shear rate (the pressure dependence is defined in equation~\eqref{Def:visc-press}), for three different values of the nonlinearity exponent $\lambda$. The left column shows results for a non-reentrant angular geometry with opening angle $\alpha = \frac{3\pi}{4}$ when $\delta = 1$: symmetric flow  (top), and antisymmetric flow (bottom). The right column displays the corresponding zoomed-in views near the corner region.}
    \label{fig:SYM-ASYM-pi3-4}
\end{figure}


\section{Results and Discussions}\label{dc}
In this section, we present and discuss the numerical results concerning the flow behavior of a fluid whose viscosity depends on both pressure and shear rate.  We used MATLAB 2022a’s ode45 solver, which provides a good balance between accuracy and efficiency.  To ascertain the results accuracy, we compared the numerical solution in benchmark cases where analytical solutions are available (i.e., Newtonian and purely piezo-viscous fluids), and the results were consistent.

Specifically, the pressure dependence of viscosity is described by equation~\eqref{Def:visc-press}. Two flow configurations are considered: symmetric and antisymmetric, each examined within angular domains characterized by an opening angle of $\alpha = \frac{3\pi}{4}$ (reentrant geometry) and $\alpha = \frac{\pi}{3}$ (non-reentrant geometry).  
The solutions of the nonlinear eigenvalue problems  \eqref{EqDiff:p_pol-4}-\eqref{BC:antisym}  and \eqref{EqDiff:p_pol-4}-\eqref{BC:sym}   are obtained upon using a shooting
technique and the value of $m$ is adjusted to satisfy the second condition in \eqref{BC:antisym} when $\theta = \alpha$ for antisymmetric flows and \eqref{BC:sym} when $\theta = \alpha$ for symmetric flows, respectively. 

The components of the flux vectors $q_u$ and $q_v$ for both antisymmetric and symmetric flow configurations are shown in Figures \ref{fig:SYM_pi3} - \ref{fig:ASYM_3pi4}, for various values of $\delta$ and $\lambda$. We recall that the non-linear exponent $n$ used in~\cite{Hassager1988-xz} is related to $\lambda$ by the relation $n = \frac{1}{\lambda}$. The classical Newtonian behavior is recovered in the limit $\delta \rightarrow 0$ when $\lambda = 1$.

Consistent with the findings reported in~\cite{Hassager1988-xz}, in the antisymmetric configuration with a reentrant angle ($\alpha = \frac{3\pi}{4}$), the flow rate increases compared to the symmetric case. This enhancement may arise from the combined influence of geometric asymmetry. 
In contrast, the non-reentrant geometry ($\alpha = \frac{\pi}{3}$) may impose greater flow restrictions due to its narrower angular confinement. 
For both angular configurations, the nonlinearity of the viscosity leads to flow characteristics that deviate markedly from the linear case (classical Newtonian or the sole piezo-viscous behavior). 
It is worth noting that the influence of pressure-dependent viscosity alone—i.e., piezo-viscosity—on the flow field is generally limited. While previous analyses (e.g.,~\cite{Calusi2024b}) suggested a more prominent role, refined simulations performed in the present work indicate that such effects are negligible in the absence of shear-rate dependence. 
We finally present on Figures \ref{fig:SYM-ASYM-pi3-delta-05} - \ref{fig:SYM-ASYM-pi3-4} the components of flux vectors $q_u$ and $q_v$ for a fluid with both pressure and shear rate dependent viscosity (see equation ~\eqref{Def:visc-press}) for which the pressure coefficient $\delta=0.5,\ 1$ by varying $\lambda$.  Both symmetric and antisymmetric flows are numerically simulated.  The results are similar in nature to the previous ones: flow rate increases in the case of $\alpha=\frac{3\pi}{4}$ compared to the case where $\alpha=\frac{\pi}{3}$ for both symmetric and antisymmetric flows. Moreover, when \(\lambda \ne 1\), especially for larger deviations from unity, the flow patterns change more significantly, especially in shear-thinning cases (\( \lambda > 1 \)). 

 Moreover, the results displayed on Figures 5 to 8 show that the pressure dependent  parameter $\delta$ varies from 0 (no piezo-viscous effect) to 1, while the rheology related parameter $\lambda
$ (it controls shear thinning or shear thickening patterns) varies from 0.83 to 1.25 (numerical values selected to exemplify).  We also observe from these figures that the pressure dependent viscosity alone exerts only a limited influence.  This is likely related to the rheological constitutive equation studied in this paper.  Different results would likely be obtained if a different rheological constitutive equation is put to work.

These findings highlight the intricate interaction between domain geometry, symmetry conditions, and viscosity dependencies, emphasizing the necessity of incorporating both pressure and shear-rate effects when modeling fluid behavior in angular domains.

Last but not least: a comparison with experimental data would of course be utterly beneficial.  However, to the best of our knowledge, no experimental data relevant to this study are currently available in the open literature.  We hope  this work  will pave the way to initiate the necessary  experimental work and stimulate  interactions between theoretical and experimental approaches.

\section{Conclusions}\label{cn}

In this work, we investigated the flow of a fluid with viscosity depending on both pressure and shear rate within a Hele-Shaw cell characterized by angular geometries. The model extends previous formulations of piezo-viscous and yield-stress fluids, aiming to provide a more comprehensive understanding of complex flow behavior near sharp corners.

A key contribution of this study is the comparison across different regimes of nonlinearity. In the visualizations (Figures~\ref{fig:SYM-ASYM-pi3-delta-05}--\ref{fig:SYM-ASYM-pi3-4}), the blue arrows represent the velocity field for $\lambda = 1.25$, while the colored arrows correspond to other combinations of $\lambda$ and $\delta$. Although none of these cases are strictly Newtonian, the configuration with $\lambda = 1$ exhibits flow structures that closely resemble Newtonian behavior, particularly as $\delta$ varies. This weak sensitivity to pressure dependence explains the relatively small variation in the component of $q_u,q_v$ for $\lambda = 1$, making it a useful reference for identifying truly nonlinear deviations. In fact, Figures \ref{fig:SYM_pi3} - \ref{fig:ASYM_3pi4} show that pressure-dependent viscosity alone exerts a limited effect. 

Our results show that when $\lambda \ne 1$, especially for higher values (shear-thinning cases), the combination of shear-rate and pressure dependence leads to pronounced differences in the flow components $(q_u,q_v)$. These effects are amplified in reentrant geometries and under antisymmetric boundary conditions, whereas non-reentrant domains tend to reduce such nonlinear responses.

These findings highlight the importance of accounting for both shear and pressure contributions when modeling confined flows in complex geometries. This insight is especially relevant to microfluidics, lubrication, and thin-film applications, where local geometric singularities can dominate the overall system behavior. Future work will investigate the stability of such flows and explore the onset of possible bifurcations and instabilities under varying nonlinearity parameters.


\begin{acknowledgments}\label{ak}

The Authors thank the anonymous Referees for their very insightful and enlightening remarks on the originally submitted manuscript. 

L. I. Palade gratefully acknowledges funding for this research was provided by the ANR-21-SFRI-0001 "Structuration de la formation par la recherche \`a l'Universit\'e de Lyon", DIGIT-BIOMED (Digital Sciences for Biology and Health) program, via an endowment awarded to him. B.C. has been funded by the NextGenerationEU PRIN 2022 research project ‘Mathematical Modelling of Heterogeneous Systems’ (grant n. 2022MKB7MM) and performed this study under the auspices of the GNFM of Italian INDAM.

\end{acknowledgments}

\nocite{*}
\bibliography{aipsamp}

\providecommand{\noopsort}[1]{}\providecommand{\singleletter}[1]{#1}%
\begin{thebibliography}{58}%
\makeatletter
\providecommand \@ifxundefined [1]{%
 \@ifx{#1\undefined}
}%
\providecommand \@ifnum [1]{%
 \ifnum #1\expandafter \@firstoftwo
 \else \expandafter \@secondoftwo
 \fi
}%
\providecommand \@ifx [1]{%
 \ifx #1\expandafter \@firstoftwo
 \else \expandafter \@secondoftwo
 \fi
}%
\providecommand \natexlab [1]{#1}%
\providecommand \enquote  [1]{``#1''}%
\providecommand \bibnamefont  [1]{#1}%
\providecommand \bibfnamefont [1]{#1}%
\providecommand \citenamefont [1]{#1}%
\providecommand \href@noop [0]{\@secondoftwo}%
\providecommand \href [0]{\begingroup \@sanitize@url \@href}%
\providecommand \@href[1]{\@@startlink{#1}\@@href}%
\providecommand \@@href[1]{\endgroup#1\@@endlink}%
\providecommand \@sanitize@url [0]{\catcode `\\12\catcode `\$12\catcode
  `\&12\catcode `\#12\catcode `\^12\catcode `\_12\catcode `\%12\relax}%
\providecommand \@@startlink[1]{}%
\providecommand \@@endlink[0]{}%
\providecommand \url  [0]{\begingroup\@sanitize@url \@url }%
\providecommand \@url [1]{\endgroup\@href {#1}{\urlprefix }}%
\providecommand \urlprefix  [0]{URL }%
\providecommand \Eprint [0]{\href }%
\providecommand \doibase [0]{http://dx.doi.org/}%
\providecommand \selectlanguage [0]{\@gobble}%
\providecommand \bibinfo  [0]{\@secondoftwo}%
\providecommand \bibfield  [0]{\@secondoftwo}%
\providecommand \translation [1]{[#1]}%
\providecommand \BibitemOpen [0]{}%
\providecommand \bibitemStop [0]{}%
\providecommand \bibitemNoStop [0]{.\EOS\space}%
\providecommand \EOS [0]{\spacefactor3000\relax}%
\providecommand \BibitemShut  [1]{\csname bibitem#1\endcsname}%
\let\auto@bib@innerbib\@empty
\bibitem [{\citenamefont {Hele-Shaw}(1898)}]{HELESHAW1898}%
  \BibitemOpen
  \bibfield  {author} {\bibinfo {author} {\bibfnamefont {H.~S.}\ \bibnamefont
  {Hele-Shaw}},\ }\bibfield  {title} {\enquote {\bibinfo {title} {{T}he {F}low
  of {W}ater},}\ }\href {\doibase 10.1038/058034a0} {\bibfield  {journal}
  {\bibinfo  {journal} {Nature}\ }\textbf {\bibinfo {volume} {58}},\ \bibinfo
  {pages} {34–36} (\bibinfo {year} {1898})}\BibitemShut {NoStop}%
\bibitem [{\citenamefont {Xu}(1998)}]{jjx1998}%
  \BibitemOpen
  \bibfield  {author} {\bibinfo {author} {\bibfnamefont {J.-J.}\ \bibnamefont
  {Xu}},\ }\href@noop {} {\emph {\bibinfo {title} {Interfacial Wave Theory of
  Pattern Formation. Selection of Dendritic Growth and Viscous Fingering in
  Hele-Shaw Flow}}}\ (\bibinfo  {publisher} {Springer-Verlag Berlin
  Heidelberg},\ \bibinfo {year} {1998})\BibitemShut {NoStop}%
\bibitem [{\citenamefont {Pop}\ and\ \citenamefont {Ingham}(2001)}]{pop2001}%
  \BibitemOpen
  \bibfield  {author} {\bibinfo {author} {\bibfnamefont {I.}~\bibnamefont
  {Pop}}\ and\ \bibinfo {author} {\bibfnamefont {D.~B.}\ \bibnamefont
  {Ingham}},\ }\href@noop {} {\emph {\bibinfo {title} {{C}onvective {H}eat
  {T}ransfer. Mathematical and Computational Modelling of {V}iscous {F}luids
  and {P}orous {M}edia}}}\ (\bibinfo  {publisher} {Elsevier},\ \bibinfo {year}
  {2001})\BibitemShut {NoStop}%
\bibitem [{\citenamefont {Nield}\ and\ \citenamefont {Bejan}(2006)}]{nb2006}%
  \BibitemOpen
  \bibfield  {author} {\bibinfo {author} {\bibfnamefont {D.~A.}\ \bibnamefont
  {Nield}}\ and\ \bibinfo {author} {\bibfnamefont {A.}~\bibnamefont {Bejan}},\
  }\href@noop {} {\emph {\bibinfo {title} {Convection in Porous Media, 3rd
  Edition}}}\ (\bibinfo  {publisher} {Springer Science+Business Media, Inc.},\
  \bibinfo {year} {2006})\BibitemShut {NoStop}%
\bibitem [{\citenamefont {Coussot}(2014)}]{coussot2014}%
  \BibitemOpen
  \bibfield  {author} {\bibinfo {author} {\bibfnamefont {P.}~\bibnamefont
  {Coussot}},\ }\href@noop {} {\emph {\bibinfo {title} {{R}heophysics. {M}atter
  in {A}ll its {S}tates.}}}\ (\bibinfo  {publisher} {Springer International
  Publishing Switzerland},\ \bibinfo {year} {2014})\BibitemShut {NoStop}%
\bibitem [{\citenamefont {Huilgol}\ and\ \citenamefont
  {Georgiou}(2015)}]{huilgeo2015}%
  \BibitemOpen
  \bibfield  {author} {\bibinfo {author} {\bibfnamefont {R.~R.}\ \bibnamefont
  {Huilgol}}\ and\ \bibinfo {author} {\bibfnamefont {G.~C.}\ \bibnamefont
  {Georgiou}},\ }\href@noop {} {\emph {\bibinfo {title} {{F}luid {M}echanics of
  {V}iscoplasticity, 2nd {E}dition}}}\ (\bibinfo  {publisher} {Springer-Verlag
  Berlin Heidelberg},\ \bibinfo {year} {2015})\BibitemShut {NoStop}%
\bibitem [{\citenamefont {Shenoy}, \citenamefont {Sheremet},\ and\
  \citenamefont {Pop}(2017)}]{pop2017}%
  \BibitemOpen
  \bibfield  {author} {\bibinfo {author} {\bibfnamefont {A.}~\bibnamefont
  {Shenoy}}, \bibinfo {author} {\bibfnamefont {M.}~\bibnamefont {Sheremet}}, \
  and\ \bibinfo {author} {\bibfnamefont {I.}~\bibnamefont {Pop}},\ }\href@noop
  {} {\emph {\bibinfo {title} {{C}onvective {F}low and {H}eat {T}ransfer {F}rom
  {W}avy {S}urfaces. {V}iscous {F}luids, {P}orous {M}edia, and {N}anofluids}}}\
  (\bibinfo  {publisher} {CRC Press, Boca Raton FL},\ \bibinfo {year}
  {2017})\BibitemShut {NoStop}%
\bibitem [{\citenamefont {Evans}\ \emph {et~al.}(2025)\citenamefont {Evans},
  \citenamefont {Palhares~Junior}, \citenamefont {Oishi},\ and\ \citenamefont
  {Ruano~Neto}}]{evans2025}%
  \BibitemOpen
  \bibfield  {author} {\bibinfo {author} {\bibfnamefont {J.~D.}\ \bibnamefont
  {Evans}}, \bibinfo {author} {\bibfnamefont {I.~C.}\ \bibnamefont
  {Palhares~Junior}}, \bibinfo {author} {\bibfnamefont {C.~M.}\ \bibnamefont
  {Oishi}}, \ and\ \bibinfo {author} {\bibfnamefont {F.}~\bibnamefont
  {Ruano~Neto}},\ }\bibfield  {title} {\enquote {\bibinfo {title} {Analysis of
  {N}ewtonian fluid flows around sharp corners with slip boundary
  conditions},}\ }\href@noop {} {\bibfield  {journal} {\bibinfo  {journal}
  {Theoretical and Computational Fluid Dynamics}\ }\textbf {\bibinfo {volume}
  {39}},\ \bibinfo {pages} {38} (\bibinfo {year} {2025})}\BibitemShut {NoStop}%
\bibitem [{\citenamefont {Hassager}\ and\ \citenamefont
  {Lauridsen}(1988)}]{Hassager1988-xz}%
  \BibitemOpen
  \bibfield  {author} {\bibinfo {author} {\bibfnamefont {O.}~\bibnamefont
  {Hassager}}\ and\ \bibinfo {author} {\bibfnamefont {T.~L.}\ \bibnamefont
  {Lauridsen}},\ }\bibfield  {title} {\enquote {\bibinfo {title} {Singular
  behavior of power - law fluids in {H}ele - {S}haw flow},}\ }\href@noop {}
  {\bibfield  {journal} {\bibinfo  {journal} {Journal of Non-Newtonian Fluid
  Mechanics}\ }\textbf {\bibinfo {volume} {42}},\ \bibinfo {pages} {337--346}
  (\bibinfo {year} {1988})}\BibitemShut {NoStop}%
\bibitem [{\citenamefont {Chupin}\ and\ \citenamefont
  {Palade}(2008)}]{Chupin2008}%
  \BibitemOpen
  \bibfield  {author} {\bibinfo {author} {\bibfnamefont {L.}~\bibnamefont
  {Chupin}}\ and\ \bibinfo {author} {\bibfnamefont {L.~I.}\ \bibnamefont
  {Palade}},\ }\bibfield  {title} {\enquote {\bibinfo {title} {Generalized
  {N}ewtonian and {H}erschel–{B}ulkley yield stress fluids pressure behavior
  near the tip of a sharp edge in thin film flows},}\ }\href {\doibase
  10.1016/j.physleta.2008.08.061} {\bibfield  {journal} {\bibinfo  {journal}
  {Physics Letters A}\ }\textbf {\bibinfo {volume} {372}},\ \bibinfo {pages}
  {6404–6411} (\bibinfo {year} {2008})}\BibitemShut {NoStop}%
\bibitem [{\citenamefont {Allouche}\ \emph {et~al.}(2015)\citenamefont
  {Allouche}, \citenamefont {Millet}, \citenamefont {Botton}, \citenamefont
  {Henry}, \citenamefont {Hadid},\ and\ \citenamefont
  {Rousset}}]{Allouche2015}%
  \BibitemOpen
  \bibfield  {author} {\bibinfo {author} {\bibfnamefont {M.~H.}\ \bibnamefont
  {Allouche}}, \bibinfo {author} {\bibfnamefont {S.}~\bibnamefont {Millet}},
  \bibinfo {author} {\bibfnamefont {V.}~\bibnamefont {Botton}}, \bibinfo
  {author} {\bibfnamefont {D.}~\bibnamefont {Henry}}, \bibinfo {author}
  {\bibfnamefont {H.~B.}\ \bibnamefont {Hadid}}, \ and\ \bibinfo {author}
  {\bibfnamefont {F.}~\bibnamefont {Rousset}},\ }\bibfield  {title} {\enquote
  {\bibinfo {title} {Stability of a flow down an incline with respect to
  two-dimensional and three-dimensional disturbances for {N}ewtonian and
  non-{N}ewtonian fluids},}\ }\href {\doibase 10.1103/physreve.92.063010}
  {\bibfield  {journal} {\bibinfo  {journal} {Physical Review E}\ }\textbf
  {\bibinfo {volume} {92}} (\bibinfo {year} {2015}),\
  10.1103/physreve.92.063010}\BibitemShut {NoStop}%
\bibitem [{\citenamefont {Allouche}\ \emph {et~al.}(2017)\citenamefont
  {Allouche}, \citenamefont {Botton}, \citenamefont {Millet}, \citenamefont
  {Henry}, \citenamefont {Dagois-Bohy}, \citenamefont {G\"{u}zel},\ and\
  \citenamefont {Hadid}}]{Allouche2017}%
  \BibitemOpen
  \bibfield  {author} {\bibinfo {author} {\bibfnamefont {M.~H.}\ \bibnamefont
  {Allouche}}, \bibinfo {author} {\bibfnamefont {V.}~\bibnamefont {Botton}},
  \bibinfo {author} {\bibfnamefont {S.}~\bibnamefont {Millet}}, \bibinfo
  {author} {\bibfnamefont {D.}~\bibnamefont {Henry}}, \bibinfo {author}
  {\bibfnamefont {S.}~\bibnamefont {Dagois-Bohy}}, \bibinfo {author}
  {\bibfnamefont {B.}~\bibnamefont {G\"{u}zel}}, \ and\ \bibinfo {author}
  {\bibfnamefont {H.~B.}\ \bibnamefont {Hadid}},\ }\bibfield  {title} {\enquote
  {\bibinfo {title} {Primary instability of a shear-thinning film flow down an
  incline: experimental study},}\ }\href {\doibase 10.1017/jfm.2017.276}
  {\bibfield  {journal} {\bibinfo  {journal} {Journal of Fluid Mechanics}\
  }\textbf {\bibinfo {volume} {821}} (\bibinfo {year} {2017}),\
  10.1017/jfm.2017.276}\BibitemShut {NoStop}%
\bibitem [{\citenamefont {Balmforth}\ and\ \citenamefont
  {Liu}(2004)}]{BALMFORTH2004}%
  \BibitemOpen
  \bibfield  {author} {\bibinfo {author} {\bibfnamefont {N.~J.}\ \bibnamefont
  {Balmforth}}\ and\ \bibinfo {author} {\bibfnamefont {J.~J.}\ \bibnamefont
  {Liu}},\ }\bibfield  {title} {\enquote {\bibinfo {title} {Roll waves in
  mud},}\ }\href {\doibase 10.1017/s0022112004000801} {\bibfield  {journal}
  {\bibinfo  {journal} {Journal of Fluid Mechanics}\ }\textbf {\bibinfo
  {volume} {519}},\ \bibinfo {pages} {33--54} (\bibinfo {year}
  {2004})}\BibitemShut {NoStop}%
\bibitem [{\citenamefont {Benjamin}(1957)}]{Benjamin1957}%
  \BibitemOpen
  \bibfield  {author} {\bibinfo {author} {\bibfnamefont {T.~B.}\ \bibnamefont
  {Benjamin}},\ }\bibfield  {title} {\enquote {\bibinfo {title} {Wave formation
  in laminar flow down an inclined plane},}\ }\href {\doibase
  10.1017/s0022112057000373} {\bibfield  {journal} {\bibinfo  {journal}
  {Journal of Fluid Mechanics}\ }\textbf {\bibinfo {volume} {2}},\ \bibinfo
  {pages} {554} (\bibinfo {year} {1957})}\BibitemShut {NoStop}%
\bibitem [{\citenamefont {Borsi}\ \emph {et~al.}(2008)\citenamefont {Borsi},
  \citenamefont {Farina}, \citenamefont {Fasano},\ and\ \citenamefont
  {Rajagopal}}]{borsi-2008}%
  \BibitemOpen
  \bibfield  {author} {\bibinfo {author} {\bibfnamefont {I.}~\bibnamefont
  {Borsi}}, \bibinfo {author} {\bibfnamefont {A.}~\bibnamefont {Farina}},
  \bibinfo {author} {\bibfnamefont {A.}~\bibnamefont {Fasano}}, \ and\ \bibinfo
  {author} {\bibfnamefont {K.~R.}\ \bibnamefont {Rajagopal}},\ }\bibfield
  {title} {\enquote {\bibinfo {title} {Modelling the combined chemical and
  mechanical action for blood clotting},}\ }\href@noop {} {\bibfield  {journal}
  {\bibinfo  {journal} {Nonlinear Phenomena with Energy Dissipation, Gakuto
  Internat Ser Math Sci Appl, Gakkotosho, Tokyo}\ }\textbf {\bibinfo {volume}
  {29}},\ \bibinfo {pages} {53--72} (\bibinfo {year} {2008})}\BibitemShut
  {NoStop}%
\bibitem [{\citenamefont {Chakraborty}, \citenamefont {Sheu},\ and\
  \citenamefont {Ghosh}(2019)}]{Chakraborty2019}%
  \BibitemOpen
  \bibfield  {author} {\bibinfo {author} {\bibfnamefont {S.}~\bibnamefont
  {Chakraborty}}, \bibinfo {author} {\bibfnamefont {T.~W.-H.}\ \bibnamefont
  {Sheu}}, \ and\ \bibinfo {author} {\bibfnamefont {S.}~\bibnamefont {Ghosh}},\
  }\bibfield  {title} {\enquote {\bibinfo {title} {Dynamics and stability of a
  power-law film flowing down a slippery slope},}\ }\href {\doibase
  10.1063/1.5078450} {\bibfield  {journal} {\bibinfo  {journal} {Physics of
  Fluids}\ }\textbf {\bibinfo {volume} {31}},\ \bibinfo {pages} {013102}
  (\bibinfo {year} {2019})}\BibitemShut {NoStop}%
\bibitem [{\citenamefont {Farina}\ and\ \citenamefont
  {Fusi}(2018)}]{Farina2018}%
  \BibitemOpen
  \bibfield  {author} {\bibinfo {author} {\bibfnamefont {A.}~\bibnamefont
  {Farina}}\ and\ \bibinfo {author} {\bibfnamefont {L.}~\bibnamefont {Fusi}},\
  }\enquote {\bibinfo {title} {Viscoplastic fluids: Mathematical modeling and
  applications},}\ in\ \href {\doibase 10.1007/978-3-319-74796-5_5} {\emph
  {\bibinfo {booktitle} {Non-Newtonian Fluid Mechanics and Complex Flows:
  Levico Terme, Italy 2016}}},\ \bibinfo {editor} {edited by\ \bibinfo {editor}
  {\bibfnamefont {A.}~\bibnamefont {Farina}}, \bibinfo {editor} {\bibfnamefont
  {A.}~\bibnamefont {Mikeli{\'{c}}}}, \ and\ \bibinfo {editor} {\bibfnamefont
  {F.}~\bibnamefont {Rosso}}}\ (\bibinfo  {publisher} {Springer International
  Publishing},\ \bibinfo {address} {Cham},\ \bibinfo {year} {2018})\ pp.\
  \bibinfo {pages} {229--298}\BibitemShut {NoStop}%
\bibitem [{\citenamefont {Gholinezhad}, \citenamefont {Kantzas},\ and\
  \citenamefont {Bryant}(2023)}]{Gholinezhad2023-dh}%
  \BibitemOpen
  \bibfield  {author} {\bibinfo {author} {\bibfnamefont {S.}~\bibnamefont
  {Gholinezhad}}, \bibinfo {author} {\bibfnamefont {A.}~\bibnamefont
  {Kantzas}}, \ and\ \bibinfo {author} {\bibfnamefont {S.~L.}\ \bibnamefont
  {Bryant}},\ }\bibfield  {title} {\enquote {\bibinfo {title} {Control of
  interfacial instabilities through variable injection rate in a radial
  hele-shaw cell: A nonlinear approach for late-time analysis},}\ }\href@noop
  {} {\bibfield  {journal} {\bibinfo  {journal} {Phys. Rev. E.}\ }\textbf
  {\bibinfo {volume} {107}},\ \bibinfo {pages} {065108} (\bibinfo {year}
  {2023})}\BibitemShut {NoStop}%
\bibitem [{\citenamefont {Hinterm\"{u}ller}\ and\ \citenamefont
  {Keil}(2021)}]{Hintermuller2021}%
  \BibitemOpen
  \bibfield  {author} {\bibinfo {author} {\bibfnamefont {M.}~\bibnamefont
  {Hinterm\"{u}ller}}\ and\ \bibinfo {author} {\bibfnamefont {T.}~\bibnamefont
  {Keil}},\ }\enquote {\bibinfo {title} {Optimal control of geometric partial
  differential equations},}\ in\ \href {\doibase 10.1016/bs.hna.2020.10.003}
  {\emph {\bibinfo {booktitle} {Geometric Partial Differential Equations - Part
  II}}}\ (\bibinfo  {publisher} {Elsevier},\ \bibinfo {year} {2021})\ p.\
  \bibinfo {pages} {213–270}\BibitemShut {NoStop}%
\bibitem [{\citenamefont {Li}, \citenamefont {Huang},\ and\ \citenamefont
  {Zhao}(2022)}]{Li2022}%
  \BibitemOpen
  \bibfield  {author} {\bibinfo {author} {\bibfnamefont {P.}~\bibnamefont
  {Li}}, \bibinfo {author} {\bibfnamefont {X.}~\bibnamefont {Huang}}, \ and\
  \bibinfo {author} {\bibfnamefont {Y.-P.}\ \bibnamefont {Zhao}},\ }\bibfield
  {title} {\enquote {\bibinfo {title} {Active control of
  electro-visco-fingering in {H}ele-{S}haw cells using {M}axwell stress},}\
  }\href {\doibase 10.1016/j.isci.2022.105204} {\bibfield  {journal} {\bibinfo
  {journal} {iScience}\ }\textbf {\bibinfo {volume} {25}},\ \bibinfo {pages}
  {105204} (\bibinfo {year} {2022})}\BibitemShut {NoStop}%
\bibitem [{\citenamefont {Taylor-West}\ and\ \citenamefont
  {Hogg}(2022)}]{taylor2022}%
  \BibitemOpen
  \bibfield  {author} {\bibinfo {author} {\bibfnamefont {J.~J.}\ \bibnamefont
  {Taylor-West}}\ and\ \bibinfo {author} {\bibfnamefont {A.~J.}\ \bibnamefont
  {Hogg}},\ }\bibfield  {title} {\enquote {\bibinfo {title} {Viscoplastic
  corner eddies},}\ }\href@noop {} {\bibfield  {journal} {\bibinfo  {journal}
  {Journal of Fluid Mechanics}\ }\textbf {\bibinfo {volume} {941}},\ \bibinfo
  {pages} {A64--1 -- A64--23} (\bibinfo {year} {2022})}\BibitemShut {NoStop}%
\bibitem [{\citenamefont {Pascal}(1999)}]{Pascal_1999}%
  \BibitemOpen
  \bibfield  {author} {\bibinfo {author} {\bibfnamefont {J.~P.}\ \bibnamefont
  {Pascal}},\ }\bibfield  {title} {\enquote {\bibinfo {title} {Linear stability
  of fluid flow down a porous inclined plane},}\ }\href {\doibase
  10.1088/0022-3727/32/4/011} {\bibfield  {journal} {\bibinfo  {journal}
  {Journal of Physics D: Applied Physics}\ }\textbf {\bibinfo {volume} {32}},\
  \bibinfo {pages} {417--422} (\bibinfo {year} {1999})}\BibitemShut {NoStop}%
\bibitem [{\citenamefont {Petit}(2024)}]{Petit2024-fi}%
  \BibitemOpen
  \bibfield  {author} {\bibinfo {author} {\bibfnamefont {N.}~\bibnamefont
  {Petit}},\ }\bibfield  {title} {\enquote {\bibinfo {title} {Optimal control
  of viscous fingering},}\ }\href@noop {} {\bibfield  {journal} {\bibinfo
  {journal} {J. Process Control}\ }\textbf {\bibinfo {volume} {135}},\ \bibinfo
  {pages} {103150} (\bibinfo {year} {2024})}\BibitemShut {NoStop}%
\bibitem [{\citenamefont {Rajagopal}, \citenamefont {Saccomandi},\ and\
  \citenamefont {Vergori}(2012)}]{Rajagopal2012}%
  \BibitemOpen
  \bibfield  {author} {\bibinfo {author} {\bibfnamefont {K.~R.}\ \bibnamefont
  {Rajagopal}}, \bibinfo {author} {\bibfnamefont {G.}~\bibnamefont
  {Saccomandi}}, \ and\ \bibinfo {author} {\bibfnamefont {L.}~\bibnamefont
  {Vergori}},\ }\bibfield  {title} {\enquote {\bibinfo {title} {Flow of fluids
  with pressure- and shear-dependent viscosity down an inclined plane},}\
  }\href {\doibase 10.1017/jfm.2012.244} {\bibfield  {journal} {\bibinfo
  {journal} {Journal of Fluid Mechanics}\ }\textbf {\bibinfo {volume} {706}},\
  \bibinfo {pages} {173–189} (\bibinfo {year} {2012})}\BibitemShut {NoStop}%
\bibitem [{\citenamefont {Falsaperla}, \citenamefont {Giacobbe},\ and\
  \citenamefont {Mulone}(2020)}]{Falsaperla2020}%
  \BibitemOpen
  \bibfield  {author} {\bibinfo {author} {\bibfnamefont {P.}~\bibnamefont
  {Falsaperla}}, \bibinfo {author} {\bibfnamefont {A.}~\bibnamefont
  {Giacobbe}}, \ and\ \bibinfo {author} {\bibfnamefont {G.}~\bibnamefont
  {Mulone}},\ }\bibfield  {title} {\enquote {\bibinfo {title} {Stability of the
  plane {B}ingham{\textendash}{P}oiseuille flow in an inclined channel},}\
  }\href {\doibase 10.3390/fluids5030141} {\bibfield  {journal} {\bibinfo
  {journal} {Fluids}\ }\textbf {\bibinfo {volume} {5}},\ \bibinfo {pages} {141}
  (\bibinfo {year} {2020})}\BibitemShut {NoStop}%
\bibitem [{\citenamefont {Fusi}(2018)}]{Fusi2018}%
  \BibitemOpen
  \bibfield  {author} {\bibinfo {author} {\bibfnamefont {L.}~\bibnamefont
  {Fusi}},\ }\bibfield  {title} {\enquote {\bibinfo {title} {Channel flow of
  viscoplastic fluids with pressure-dependent rheological parameters},}\ }\href
  {\doibase 10.1063/1.5042330} {\bibfield  {journal} {\bibinfo  {journal}
  {Physics of Fluids}\ }\textbf {\bibinfo {volume} {30}},\ \bibinfo {pages}
  {073102} (\bibinfo {year} {2018})}\BibitemShut {NoStop}%
\bibitem [{\citenamefont {Huilgol}(2006)}]{Huilgol2006}%
  \BibitemOpen
  \bibfield  {author} {\bibinfo {author} {\bibfnamefont {R.}~\bibnamefont
  {Huilgol}},\ }\bibfield  {title} {\enquote {\bibinfo {title} {On the
  derivation of the symmetric and asymmetric {H}ele–{S}haw flow equations for
  viscous and viscoplastic fluids using the viscometric fluidity function},}\
  }\href {\doibase 10.1016/j.jnnfm.2006.07.008} {\bibfield  {journal} {\bibinfo
   {journal} {Journal of Non-Newtonian Fluid Mechanics}\ }\textbf {\bibinfo
  {volume} {138}},\ \bibinfo {pages} {209–213} (\bibinfo {year}
  {2006})}\BibitemShut {NoStop}%
\bibitem [{\citenamefont {Kislaya}\ \emph {et~al.}(2025)\citenamefont
  {Kislaya}, \citenamefont {Samant}, \citenamefont {Veenstra}, \citenamefont
  {Tam},\ and\ \citenamefont {Westerweel}}]{Kislaya2025}%
  \BibitemOpen
  \bibfield  {author} {\bibinfo {author} {\bibfnamefont {A.}~\bibnamefont
  {Kislaya}}, \bibinfo {author} {\bibfnamefont {A.~A.}\ \bibnamefont {Samant}},
  \bibinfo {author} {\bibfnamefont {P.}~\bibnamefont {Veenstra}}, \bibinfo
  {author} {\bibfnamefont {D.~S.~W.}\ \bibnamefont {Tam}}, \ and\ \bibinfo
  {author} {\bibfnamefont {J.}~\bibnamefont {Westerweel}},\ }\bibfield  {title}
  {\enquote {\bibinfo {title} {Particle manipulation in {H}ele–{S}haw flow
  with programmable hydrodynamics},}\ }\href {\doibase 10.1063/5.0251563}
  {\bibfield  {journal} {\bibinfo  {journal} {Physics of Fluids}\ }\textbf
  {\bibinfo {volume} {37}},\ \bibinfo {pages} {032012} (\bibinfo {year}
  {2025})}\BibitemShut {NoStop}%
\bibitem [{\citenamefont {Del~Mastro}\ and\ \citenamefont
  {Terzis}(2025)}]{Delmastro2025}%
  \BibitemOpen
  \bibfield  {author} {\bibinfo {author} {\bibfnamefont {M.}~\bibnamefont
  {Del~Mastro}}\ and\ \bibinfo {author} {\bibfnamefont {A.}~\bibnamefont
  {Terzis}},\ }\bibfield  {title} {\enquote {\bibinfo {title} {On the exact
  solutions of {D}arcy–{B}rinkman model in rectangular {H}ele–{S}haw
  channels under no-slip and slip boundary conditions},}\ }\href {\doibase
  10.1063/5.0256304} {\bibfield  {journal} {\bibinfo  {journal} {Physics of
  Fluids}\ }\textbf {\bibinfo {volume} {37}},\ \bibinfo {pages} {032012}
  (\bibinfo {year} {2025})}\BibitemShut {NoStop}%
\bibitem [{\citenamefont {Krakov}, \citenamefont {Chernyshov},\ and\
  \citenamefont {Zakinyan}(2025)}]{Krakov2025}%
  \BibitemOpen
  \bibfield  {author} {\bibinfo {author} {\bibfnamefont {M.~S.}\ \bibnamefont
  {Krakov}}, \bibinfo {author} {\bibfnamefont {A.~V.}\ \bibnamefont
  {Chernyshov}}, \ and\ \bibinfo {author} {\bibfnamefont {A.~R.}\ \bibnamefont
  {Zakinyan}},\ }\bibfield  {title} {\enquote {\bibinfo {title} {Fall and
  breakup of miscible magnetic fluid drops in a {H}ele–{S}haw cell},}\ }\href
  {\doibase 10.1063/5.0256956} {\bibfield  {journal} {\bibinfo  {journal}
  {Physics of Fluids}\ }\textbf {\bibinfo {volume} {37}},\ \bibinfo {pages}
  {033301} (\bibinfo {year} {2025})}\BibitemShut {NoStop}%
\bibitem [{\citenamefont {Zahid}\ \emph {et~al.}(2025)\citenamefont {Zahid},
  \citenamefont {Halkarni}, \citenamefont {Das}, \citenamefont {Hota},\ and\
  \citenamefont {Goyal}}]{Zahid2025}%
  \BibitemOpen
  \bibfield  {author} {\bibinfo {author} {\bibfnamefont {S.}~\bibnamefont
  {Zahid}}, \bibinfo {author} {\bibfnamefont {S.~S.}\ \bibnamefont {Halkarni}},
  \bibinfo {author} {\bibfnamefont {P.}~\bibnamefont {Das}}, \bibinfo {author}
  {\bibfnamefont {T.~K.}\ \bibnamefont {Hota}}, \ and\ \bibinfo {author}
  {\bibfnamefont {D.}~\bibnamefont {Goyal}},\ }\bibfield  {title} {\enquote
  {\bibinfo {title} {Effect of sinusoidal injection velocity on miscible
  thermo-viscous fingering in a rectilinear {H}ele-{S}haw cell},}\ }\href
  {\doibase 10.1063/5.0251203} {\bibfield  {journal} {\bibinfo  {journal}
  {Physics of Fluids}\ }\textbf {\bibinfo {volume} {37}},\ \bibinfo {pages}
  {024120} (\bibinfo {year} {2025})}\BibitemShut {NoStop}%
\bibitem [{\citenamefont {Firoozi}, \citenamefont {Ahmadpour},\ and\
  \citenamefont {Hajmohammadi}(2025)}]{Firoozi2025}%
  \BibitemOpen
  \bibfield  {author} {\bibinfo {author} {\bibfnamefont {A.}~\bibnamefont
  {Firoozi}}, \bibinfo {author} {\bibfnamefont {A.}~\bibnamefont {Ahmadpour}},
  \ and\ \bibinfo {author} {\bibfnamefont {M.~R.}\ \bibnamefont
  {Hajmohammadi}},\ }\bibfield  {title} {\enquote {\bibinfo {title} {Polymer
  flooding characteristics in modified {H}ele–{S}haw cells},}\ }\href
  {\doibase 10.1063/5.0250369} {\bibfield  {journal} {\bibinfo  {journal}
  {Physics of Fluids}\ }\textbf {\bibinfo {volume} {37}},\ \bibinfo {pages}
  {023316} (\bibinfo {year} {2025})}\BibitemShut {NoStop}%
\bibitem [{\citenamefont {Del~Mastro}\ and\ \citenamefont
  {Terzis}(2024)}]{Delmastro2024}%
  \BibitemOpen
  \bibfield  {author} {\bibinfo {author} {\bibfnamefont {M.}~\bibnamefont
  {Del~Mastro}}\ and\ \bibinfo {author} {\bibfnamefont {A.}~\bibnamefont
  {Terzis}},\ }\bibfield  {title} {\enquote {\bibinfo {title} {An experimental
  investigation of boundary layer over permeable interfaces in {H}ele-{S}haw
  micromodels},}\ }\href {\doibase 10.1063/5.0238046} {\bibfield  {journal}
  {\bibinfo  {journal} {Physics of Fluids}\ }\textbf {\bibinfo {volume} {36}},\
  \bibinfo {pages} {112110} (\bibinfo {year} {2024})}\BibitemShut {NoStop}%
\bibitem [{\citenamefont {Daripa}\ and\ \citenamefont
  {Gin}(2024)}]{Daripa2024}%
  \BibitemOpen
  \bibfield  {author} {\bibinfo {author} {\bibfnamefont {P.}~\bibnamefont
  {Daripa}}\ and\ \bibinfo {author} {\bibfnamefont {C.}~\bibnamefont {Gin}},\
  }\bibfield  {title} {\enquote {\bibinfo {title} {New results on the motion of
  interfaces of multi-layer radial {H}ele-{S}haw flows},}\ }\href {\doibase
  10.1063/5.0234519} {\bibfield  {journal} {\bibinfo  {journal} {Physics of
  Fluids}\ }\textbf {\bibinfo {volume} {36}},\ \bibinfo {pages} {102133}
  (\bibinfo {year} {2024})}\BibitemShut {NoStop}%
\bibitem [{\citenamefont {Ouyang}\ \emph {et~al.}(2024)\citenamefont {Ouyang},
  \citenamefont {Md~Basir}, \citenamefont {Naganthran},\ and\ \citenamefont
  {Pop}}]{Ouyang2024}%
  \BibitemOpen
  \bibfield  {author} {\bibinfo {author} {\bibfnamefont {Y.}~\bibnamefont
  {Ouyang}}, \bibinfo {author} {\bibfnamefont {M.~F.}\ \bibnamefont
  {Md~Basir}}, \bibinfo {author} {\bibfnamefont {K.}~\bibnamefont
  {Naganthran}}, \ and\ \bibinfo {author} {\bibfnamefont {I.}~\bibnamefont
  {Pop}},\ }\bibfield  {title} {\enquote {\bibinfo {title} {Unsteady
  magnetohydrodynamic tri-hybrid nanofluid flow past a moving wedge with
  viscous dissipation and {J}oule heating},}\ }\href {\doibase
  10.1063/5.0208608} {\bibfield  {journal} {\bibinfo  {journal} {Physics of
  Fluids}\ }\textbf {\bibinfo {volume} {36}} (\bibinfo {year} {2024}),\
  10.1063/5.0208608}\BibitemShut {NoStop}%
\bibitem [{\citenamefont {Calusi}\ and\ \citenamefont
  {Palade}(2024)}]{Calusi2024b}%
  \BibitemOpen
  \bibfield  {author} {\bibinfo {author} {\bibfnamefont {B.}~\bibnamefont
  {Calusi}}\ and\ \bibinfo {author} {\bibfnamefont {L.~I.}\ \bibnamefont
  {Palade}},\ }\bibfield  {title} {\enquote {\bibinfo {title} {{M}odeling of a
  {F}luid with {P}ressure-{D}ependent {V}iscosity in {H}ele-{S}haw {F}low},}\
  }\href {\doibase 10.3390/modelling5040077} {\bibfield  {journal} {\bibinfo
  {journal} {Modelling}\ }\textbf {\bibinfo {volume} {5}},\ \bibinfo {pages}
  {1490--1504} (\bibinfo {year} {2024})}\BibitemShut {NoStop}%
\bibitem [{\citenamefont {Barus}(1893)}]{Barus1893}%
  \BibitemOpen
  \bibfield  {author} {\bibinfo {author} {\bibfnamefont {C.}~\bibnamefont
  {Barus}},\ }\bibfield  {title} {\enquote {\bibinfo {title} {Isothermals,
  isopiestics and isometrics relative to viscosity},}\ }\href {\doibase
  10.2475/ajs.s3-45.266.87} {\bibfield  {journal} {\bibinfo  {journal}
  {American Journal of Science}\ }\textbf {\bibinfo {volume} {s3-45}},\
  \bibinfo {pages} {87–96} (\bibinfo {year} {1893})}\BibitemShut {NoStop}%
\bibitem [{\citenamefont {Henriksen}\ and\ \citenamefont
  {Hassager}(1989)}]{has1989}%
  \BibitemOpen
  \bibfield  {author} {\bibinfo {author} {\bibfnamefont {P.}~\bibnamefont
  {Henriksen}}\ and\ \bibinfo {author} {\bibfnamefont {O.}~\bibnamefont
  {Hassager}},\ }\bibfield  {title} {\enquote {\bibinfo {title} {{C}orner
  {F}low of {P}ower {L}aw {F}luids},}\ }\href {\doibase 10.1122/1.550039}
  {\bibfield  {journal} {\bibinfo  {journal} {Journal of Rheology}\ }\textbf
  {\bibinfo {volume} {33}},\ \bibinfo {pages} {865--879} (\bibinfo {year}
  {1989})}\BibitemShut {NoStop}%
\bibitem [{\citenamefont {Aronsson}\ and\ \citenamefont
  {Janfalk}(1992)}]{Aronsson1992}%
  \BibitemOpen
  \bibfield  {author} {\bibinfo {author} {\bibfnamefont {G.}~\bibnamefont
  {Aronsson}}\ and\ \bibinfo {author} {\bibfnamefont {U.}~\bibnamefont
  {Janfalk}},\ }\bibfield  {title} {\enquote {\bibinfo {title} {On
  {H}ele–{S}haw flow of power-law fluids},}\ }\href {\doibase
  10.1017/s0956792500000905} {\bibfield  {journal} {\bibinfo  {journal}
  {European Journal of Applied Mathematics}\ }\textbf {\bibinfo {volume} {3}},\
  \bibinfo {pages} {343–366} (\bibinfo {year} {1992})}\BibitemShut {NoStop}%
\bibitem [{\citenamefont {Hron}, \citenamefont {Málek},\ and\ \citenamefont
  {Rajagopal}(2001)}]{Hron2001}%
  \BibitemOpen
  \bibfield  {author} {\bibinfo {author} {\bibfnamefont {J.}~\bibnamefont
  {Hron}}, \bibinfo {author} {\bibfnamefont {J.}~\bibnamefont {Málek}}, \ and\
  \bibinfo {author} {\bibfnamefont {K.~R.}\ \bibnamefont {Rajagopal}},\
  }\bibfield  {title} {\enquote {\bibinfo {title} {Simple flows of fluids with
  pressure–dependent viscosities},}\ }\href {\doibase 10.1098/rspa.2000.0723}
  {\bibfield  {journal} {\bibinfo  {journal} {Proceedings of the Royal Society
  of London. Series A: Mathematical, Physical and Engineering Sciences}\
  }\textbf {\bibinfo {volume} {457}},\ \bibinfo {pages} {1603–1622} (\bibinfo
  {year} {2001})}\BibitemShut {NoStop}%
\bibitem [{\citenamefont {Rajagopal}, \citenamefont {Saccomandi},\ and\
  \citenamefont {Vergori}(2013)}]{Rajagopal2013}%
  \BibitemOpen
  \bibfield  {author} {\bibinfo {author} {\bibfnamefont {K.}~\bibnamefont
  {Rajagopal}}, \bibinfo {author} {\bibfnamefont {G.}~\bibnamefont
  {Saccomandi}}, \ and\ \bibinfo {author} {\bibfnamefont {L.}~\bibnamefont
  {Vergori}},\ }\bibfield  {title} {\enquote {\bibinfo {title} {Unsteady flows
  of fluids with pressure dependent viscosity},}\ }\href {\doibase
  10.1016/j.jmaa.2013.03.025} {\bibfield  {journal} {\bibinfo  {journal}
  {Journal of Mathematical Analysis and Applications}\ }\textbf {\bibinfo
  {volume} {404}},\ \bibinfo {pages} {362–372} (\bibinfo {year}
  {2013})}\BibitemShut {NoStop}%
\bibitem [{\citenamefont {Hieber}\ and\ \citenamefont
  {Shen}(1980)}]{Hieber1980}%
  \BibitemOpen
  \bibfield  {author} {\bibinfo {author} {\bibfnamefont {C.}~\bibnamefont
  {Hieber}}\ and\ \bibinfo {author} {\bibfnamefont {S.}~\bibnamefont {Shen}},\
  }\bibfield  {title} {\enquote {\bibinfo {title} {A
  finite-element/finite-difference simulation of the injection-molding filling
  process},}\ }\href {\doibase 10.1016/0377-0257(80)85012-9} {\bibfield
  {journal} {\bibinfo  {journal} {Journal of Non-Newtonian Fluid Mechanics}\
  }\textbf {\bibinfo {volume} {7}},\ \bibinfo {pages} {1–32} (\bibinfo {year}
  {1980})}\BibitemShut {NoStop}%
\bibitem [{\citenamefont {Bridgman}(1931)}]{Bridgman1931}%
  \BibitemOpen
  \bibfield  {author} {\bibinfo {author} {\bibfnamefont {P.}~\bibnamefont
  {Bridgman}},\ }\href@noop {} {\emph {\bibinfo {title} {{T}he {P}hysics of
  {H}igh {P}ressure}}}\ (\bibinfo  {publisher} {The Macmillan Compan, New
  York},\ \bibinfo {year} {1931})\BibitemShut {NoStop}%
\bibitem [{\citenamefont {Saccomandi}\ and\ \citenamefont
  {Vergori}(2010)}]{Saccomandi2010}%
  \BibitemOpen
  \bibfield  {author} {\bibinfo {author} {\bibfnamefont {G.}~\bibnamefont
  {Saccomandi}}\ and\ \bibinfo {author} {\bibfnamefont {L.}~\bibnamefont
  {Vergori}},\ }\bibfield  {title} {\enquote {\bibinfo {title} {Piezo-viscous
  flows over an inclined surface},}\ }\href
  {http://www.jstor.org/stable/43638955} {\bibfield  {journal} {\bibinfo
  {journal} {Quarterly of Applied Mathematics}\ }\textbf {\bibinfo {volume}
  {68}},\ \bibinfo {pages} {747–763} (\bibinfo {year} {2010})}\BibitemShut
  {NoStop}%
\bibitem [{\citenamefont {Rajagopal}(2011)}]{Rajagopal2011}%
  \BibitemOpen
  \bibfield  {author} {\bibinfo {author} {\bibfnamefont {K.~R.}\ \bibnamefont
  {Rajagopal}},\ }\bibfield  {title} {\enquote {\bibinfo {title} {Implicit
  constitutive relations},}\ }\href@noop {} {\bibfield  {journal} {\bibinfo
  {journal} {Continuum Mechanics}\ }\textbf {\bibinfo {volume} {III}} (\bibinfo
  {year} {2011})}\BibitemShut {NoStop}%
\bibitem [{\citenamefont {Fusi}\ and\ \citenamefont {Tozzi}(2024)}]{Fusi2024}%
  \BibitemOpen
  \bibfield  {author} {\bibinfo {author} {\bibfnamefont {L.}~\bibnamefont
  {Fusi}}\ and\ \bibinfo {author} {\bibfnamefont {R.}~\bibnamefont {Tozzi}},\
  }\bibfield  {title} {\enquote {\bibinfo {title} {{F}alkner–{S}kan boundary
  layer flow of a fluid with pressure-dependent viscosity past a stretching
  wedge with suction or injection},}\ }\href {\doibase
  10.1016/j.ijnonlinmec.2024.104746} {\bibfield  {journal} {\bibinfo  {journal}
  {International Journal of Non-Linear Mechanics}\ }\textbf {\bibinfo {volume}
  {163}},\ \bibinfo {pages} {104746} (\bibinfo {year} {2024})}\BibitemShut
  {NoStop}%
\bibitem [{\citenamefont {Rajagopal}(2003)}]{Rajagopal2003}%
  \BibitemOpen
  \bibfield  {author} {\bibinfo {author} {\bibfnamefont {K.~R.}\ \bibnamefont
  {Rajagopal}},\ }\bibfield  {title} {\enquote {\bibinfo {title} {On implicit
  constitutive theories},}\ }\href {\doibase 10.1023/a:1026062615145}
  {\bibfield  {journal} {\bibinfo  {journal} {Applications of Mathematics}\
  }\textbf {\bibinfo {volume} {48}},\ \bibinfo {pages} {279–319} (\bibinfo
  {year} {2003})}\BibitemShut {NoStop}%
\bibitem [{\citenamefont {Rajagopal}\ and\ \citenamefont
  {Srinivasa}(2007)}]{Rajagopal2007}%
  \BibitemOpen
  \bibfield  {author} {\bibinfo {author} {\bibfnamefont {K.~R.}\ \bibnamefont
  {Rajagopal}}\ and\ \bibinfo {author} {\bibfnamefont {A.~R.}\ \bibnamefont
  {Srinivasa}},\ }\bibfield  {title} {\enquote {\bibinfo {title} {On the
  thermodynamics of fluids defined by implicit constitutive relations},}\
  }\href {\doibase 10.1007/s00033-007-7039-1} {\bibfield  {journal} {\bibinfo
  {journal} {Zeitschrift f\"{u}r angewandte Mathematik und Physik}\ }\textbf
  {\bibinfo {volume} {59}},\ \bibinfo {pages} {715–729} (\bibinfo {year}
  {2007})}\BibitemShut {NoStop}%
\bibitem [{\citenamefont {Rajagopal}, \citenamefont {Saccomandi},\ and\
  \citenamefont {Vergori}(2009{\natexlab{a}})}]{Rajagopal2009}%
  \BibitemOpen
  \bibfield  {author} {\bibinfo {author} {\bibfnamefont {K.~R.}\ \bibnamefont
  {Rajagopal}}, \bibinfo {author} {\bibfnamefont {G.}~\bibnamefont
  {Saccomandi}}, \ and\ \bibinfo {author} {\bibfnamefont {L.}~\bibnamefont
  {Vergori}},\ }\bibfield  {title} {\enquote {\bibinfo {title} {Stability
  analysis of the {R}ayleigh–{B}énard convection for a fluid with
  temperature and pressure dependent viscosity},}\ }\href {\doibase
  10.1007/s00033-008-8062-6} {\bibfield  {journal} {\bibinfo  {journal}
  {Zeitschrift f\"{u}r angewandte Mathematik und Physik}\ }\textbf {\bibinfo
  {volume} {60}},\ \bibinfo {pages} {739–755} (\bibinfo {year}
  {2009}{\natexlab{a}})}\BibitemShut {NoStop}%
\bibitem [{\citenamefont {Rajagopal}, \citenamefont {Saccomandi},\ and\
  \citenamefont {Vergori}(2009{\natexlab{b}})}]{Rajagopal2009b}%
  \BibitemOpen
  \bibfield  {author} {\bibinfo {author} {\bibfnamefont {K.}~\bibnamefont
  {Rajagopal}}, \bibinfo {author} {\bibfnamefont {G.}~\bibnamefont
  {Saccomandi}}, \ and\ \bibinfo {author} {\bibfnamefont {L.}~\bibnamefont
  {Vergori}},\ }\bibfield  {title} {\enquote {\bibinfo {title} {On the
  {O}berbeck–{B}oussinesq approximation for fluids with pressure dependent
  viscosities},}\ }\href {\doibase 10.1016/j.nonrwa.2007.12.003} {\bibfield
  {journal} {\bibinfo  {journal} {Nonlinear Analysis: Real World Applications}\
  }\textbf {\bibinfo {volume} {10}},\ \bibinfo {pages} {1139–1150} (\bibinfo
  {year} {2009}{\natexlab{b}})}\BibitemShut {NoStop}%
\bibitem [{\citenamefont {Kondic}, \citenamefont {Palffy-Muhoray},\ and\
  \citenamefont {Shelley}(1996)}]{Kondic1996}%
  \BibitemOpen
  \bibfield  {author} {\bibinfo {author} {\bibfnamefont {L.}~\bibnamefont
  {Kondic}}, \bibinfo {author} {\bibfnamefont {P.}~\bibnamefont
  {Palffy-Muhoray}}, \ and\ \bibinfo {author} {\bibfnamefont {M.~J.}\
  \bibnamefont {Shelley}},\ }\bibfield  {title} {\enquote {\bibinfo {title}
  {Models of non-{N}ewtonian {H}ele-{S}haw flow},}\ }\href {\doibase
  10.1103/physreve.54.r4536} {\bibfield  {journal} {\bibinfo  {journal}
  {Physical Review E}\ }\textbf {\bibinfo {volume} {54}},\ \bibinfo {pages}
  {R4536–R4539} (\bibinfo {year} {1996})}\BibitemShut {NoStop}%
\bibitem [{\citenamefont {Nassehi}(1996)}]{NASSEHI1996}%
  \BibitemOpen
  \bibfield  {author} {\bibinfo {author} {\bibfnamefont {V.}~\bibnamefont
  {Nassehi}},\ }\bibfield  {title} {\enquote {\bibinfo {title} {Generalized
  {H}ele-{S}haw models for non-{N}ewtonian, nonisothermal flow in thin curved
  layers},}\ }\href {\doibase 10.1093/imaman/7.1.71} {\bibfield  {journal}
  {\bibinfo  {journal} {IMA Journal of Management Mathematics}\ }\textbf
  {\bibinfo {volume} {7}},\ \bibinfo {pages} {71–88} (\bibinfo {year}
  {1996})}\BibitemShut {NoStop}%
\bibitem [{\citenamefont {Klettner}, \citenamefont {Dang},\ and\ \citenamefont
  {Smith}(2023)}]{Klettner2023}%
  \BibitemOpen
  \bibfield  {author} {\bibinfo {author} {\bibfnamefont {C.}~\bibnamefont
  {Klettner}}, \bibinfo {author} {\bibfnamefont {T.}~\bibnamefont {Dang}}, \
  and\ \bibinfo {author} {\bibfnamefont {F.}~\bibnamefont {Smith}},\ }\bibfield
   {title} {\enquote {\bibinfo {title} {On the flow past ellipses in a
  {H}ele-{S}haw cell},}\ }\href {\doibase 10.1017/jfm.2023.527} {\bibfield
  {journal} {\bibinfo  {journal} {Journal of Fluid Mechanics}\ }\textbf
  {\bibinfo {volume} {971}} (\bibinfo {year} {2023}),\
  10.1017/jfm.2023.527}\BibitemShut {NoStop}%
\bibitem [{\citenamefont {Calusi}, \citenamefont {Fusi},\ and\ \citenamefont
  {Farina}(2015)}]{Calusi2015}%
  \BibitemOpen
  \bibfield  {author} {\bibinfo {author} {\bibfnamefont {B.}~\bibnamefont
  {Calusi}}, \bibinfo {author} {\bibfnamefont {L.}~\bibnamefont {Fusi}}, \ and\
  \bibinfo {author} {\bibfnamefont {A.}~\bibnamefont {Farina}},\ }\bibfield
  {title} {\enquote {\bibinfo {title} {On a free boundary problem arising in
  snow avalanche dynamics},}\ }\href {\doibase 10.1002/zamm.201400250}
  {\bibfield  {journal} {\bibinfo  {journal} {{ZAMM} - Journal of Applied
  Mathematics and Mechanics / Zeitschrift f\"{u}r Angewandte Mathematik und
  Mechanik}\ }\textbf {\bibinfo {volume} {96}},\ \bibinfo {pages} {453--465}
  (\bibinfo {year} {2015})}\BibitemShut {NoStop}%
\bibitem [{\citenamefont {Calusi}\ \emph {et~al.}(2024)\citenamefont {Calusi},
  \citenamefont {Farina}, \citenamefont {Fusi},\ and\ \citenamefont
  {Rosso}}]{Calusi2024}%
  \BibitemOpen
  \bibfield  {author} {\bibinfo {author} {\bibfnamefont {B.}~\bibnamefont
  {Calusi}}, \bibinfo {author} {\bibfnamefont {A.}~\bibnamefont {Farina}},
  \bibinfo {author} {\bibfnamefont {L.}~\bibnamefont {Fusi}}, \ and\ \bibinfo
  {author} {\bibfnamefont {F.}~\bibnamefont {Rosso}},\ }\bibfield  {title}
  {\enquote {\bibinfo {title} {Thermo-mechanical modeling of pancakelike domes
  on {V}enus},}\ }\href {\doibase 10.1063/5.0209674} {\bibfield  {journal}
  {\bibinfo  {journal} {Physics of Fluids}\ }\textbf {\bibinfo {volume} {36}}
  (\bibinfo {year} {2024}),\ 10.1063/5.0209674}\BibitemShut {NoStop}%
\bibitem [{\citenamefont {Calusi}\ \emph {et~al.}(2022)\citenamefont {Calusi},
  \citenamefont {Farina}, \citenamefont {Fusi},\ and\ \citenamefont
  {Palade}}]{Calusi2023}%
  \BibitemOpen
  \bibfield  {author} {\bibinfo {author} {\bibfnamefont {B.}~\bibnamefont
  {Calusi}}, \bibinfo {author} {\bibfnamefont {A.}~\bibnamefont {Farina}},
  \bibinfo {author} {\bibfnamefont {L.}~\bibnamefont {Fusi}}, \ and\ \bibinfo
  {author} {\bibfnamefont {L.~I.}\ \bibnamefont {Palade}},\ }\bibfield  {title}
  {\enquote {\bibinfo {title} {Stability of a regularized {C}asson flow down an
  incline: Comparison with the {B}ingham case},}\ }\href {\doibase
  10.3390/fluids7120380} {\bibfield  {journal} {\bibinfo  {journal} {Fluids}\
  }\textbf {\bibinfo {volume} {7}},\ \bibinfo {pages} {380} (\bibinfo {year}
  {2022})}\BibitemShut {NoStop}%
\bibitem [{\citenamefont {Calusi}, \citenamefont {Fusi},\ and\ \citenamefont
  {Farina}(2023)}]{Calusi2023b}%
  \BibitemOpen
  \bibfield  {author} {\bibinfo {author} {\bibfnamefont {B.}~\bibnamefont
  {Calusi}}, \bibinfo {author} {\bibfnamefont {L.}~\bibnamefont {Fusi}}, \ and\
  \bibinfo {author} {\bibfnamefont {A.}~\bibnamefont {Farina}},\ }\bibfield
  {title} {\enquote {\bibinfo {title} {Linear stability of a {C}ouette flow for
  non-monotone stress-power law models},}\ }\href {\doibase
  10.1140/epjp/s13360-023-04566-1} {\bibfield  {journal} {\bibinfo  {journal}
  {The European Physical Journal Plus}\ }\textbf {\bibinfo {volume} {138}}
  (\bibinfo {year} {2023}),\ 10.1140/epjp/s13360-023-04566-1}\BibitemShut
  {NoStop}%
\bibitem [{\citenamefont {Fern{\'{a}}ndez-Nieto}, \citenamefont {Noble},\ and\
  \citenamefont {Vila}(2010)}]{FernndezNieto2010}%
  \BibitemOpen
  \bibfield  {author} {\bibinfo {author} {\bibfnamefont {E.~D.}\ \bibnamefont
  {Fern{\'{a}}ndez-Nieto}}, \bibinfo {author} {\bibfnamefont {P.}~\bibnamefont
  {Noble}}, \ and\ \bibinfo {author} {\bibfnamefont {J.-P.}\ \bibnamefont
  {Vila}},\ }\bibfield  {title} {\enquote {\bibinfo {title} {Shallow water
  equations for non-{N}ewtonian fluids},}\ }\href {\doibase
  10.1016/j.jnnfm.2010.03.008} {\bibfield  {journal} {\bibinfo  {journal}
  {Journal of Non-Newtonian Fluid Mechanics}\ }\textbf {\bibinfo {volume}
  {165}},\ \bibinfo {pages} {712--732} (\bibinfo {year} {2010})}\BibitemShut
  {NoStop}%
\end{thebibliography}%

\end{document}